\documentclass[iop,apj,twocolumn]{emulateapj}
\usepackage{times}
\usepackage{natbib}          
\newcommand{\chandra}{\textit{Chandra}}
\newcommand{\spitzer}{\textit{Spitzer}}
\newcommand{\rosat}{{ROSAT}}

\newcommand{\source}{{NGC\,7016}}
\newcommand{\sseven}{{NGC\,7017}}
\newcommand{\seight}{{NGC\,7018}}

\newcommand{\cluster}{{Abell\,3744}}

\begin{document}

\slugcomment{December 2013, accepted for publication in Ap J}

\shorttitle{Sliding not sloshing in Abell 3744 }

\shortauthors{D.M.~Worrall \& M.~Birkinshaw}

\title{Sliding not sloshing in Abell 3744: the influence of radio
  galaxies NGC 7018 and 7016 on cluster gas }

\author{
D.M.~Worrall\altaffilmark{1,2}, 
M.~Birkinshaw\altaffilmark{1,2}
}
\altaffiltext{1}{HH Wills Physics Laboratory,
University of Bristol, 
Tyndall Avenue, 
Bristol\ \ BS8 1TL, UK}
\altaffiltext{2}{Harvard-Smithsonian Center for Astrophysics, 
60 Garden Street, 
Cambridge, MA\ \ 02138}
\email{d.worrall@bristol.ac.uk}

\begin{abstract}

We present new X-ray (\chandra) and radio (JVLA) observations of the
nearby cluster Abell 3744.  It hosts two prominent radio galaxies with
powers in the range critical for radio-mode feedback. The radio
emission from these galaxies terminates in buoyant tendrils reaching
the cluster's outer edge, and the radio-emitting plasma clearly
influences the cluster's X-ray-emitting atmosphere.  The cluster's
average gas temperature, of $kT=3.5$ keV, is high for its bolometric
luminosity of 3.2 $\times 10^{43}$ ergs s$^{-1}$, but the 100
kpc-scale cavity carved out by radio-emitting plasma shows evidence of
less than 2 per cent of the excess enthalpy. We suggest instead that a
high-velocity encounter with a galaxy group is responsible for
dispersing and increasing the entropy of the gas in this non-cool-core
cluster.  We see no evidence for shocks, or established isobaric gas
motions (sloshing), but there is much sub-structure associated with a
dynamically active central region that encompasses the brightest radio
emission.  Gas heating is evident in directions perpendicular to the
inferred line of encounter between the infalling group and
cluster. The radio-emitting tendrils run along boundaries between gas
of different temperature, apparently lubricating the gas flows and
inhibiting heat transfer.  The first stages of the encounter may have
helped trigger the radio galaxies into their current phase of
activity, where we see X-rays from the nuclei, jets, and hotspots.

\end{abstract}

\keywords{
galaxies: active ---
galaxies: individual (\objectname{\source, NGC 7017, \seight}) ---
galaxies: clusters: individual (\objectname{\cluster}) ---
galaxies: jets ---
radio continuum: galaxies ---
X-rays: galaxies: clusters}

\section{Introduction}
\label{sec:intro}

It is many years since the first description of how radio sources
fueled by AGN interact with and inject energy into the surrounding
medium \citep{scheuer}. Now, largely due to \chandra, we have many
observational examples.  X-ray signatures of the mechanisms involved
are varied.  Gas cavities crafted by current or past radio lobes are
common \citep[see][for a review]{mcnamara}.  AGN-driven radio lobes
are seen to shock the gas strongly, as in Cen\,A and PKS~B2152-699
\citep{kraft,wfos} or weakly, as in NGC\,4636 \citep{jones}.  While
these phenomena mold the gas surface-brightness distribution, in other
interactions it is the gas structures that shape the distribution of
radio emission.  This may be through the radio structures becoming
buoyant, as in NGC\,326 \citep*{wbc326} or M\,87 \citep{forman}, or
because radio plasma is riding on a pressure wave of gas, as in
3C\,442A \citep{w442a}.  Moreover, as galaxies within groups and
clusters interact with one another, their interstellar and
intracluster media (ISM and ICM) get churned up through ram-pressure
stripping, and wakes are observed as X-ray-gas density enhancements
which are sometimes cooler and sometimes hotter than the surrounding
gas \citep[e.g.,][]{machacek, sakelliou, randall}.

\begin{figure}
\centering
\includegraphics[width=0.9\columnwidth,clip=true]{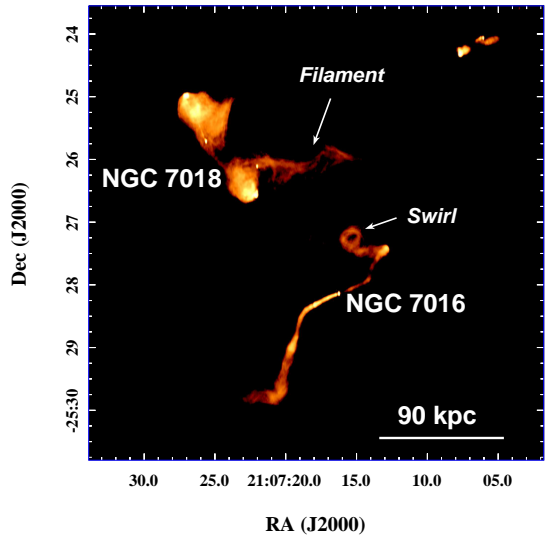}
\caption[]{\source\ and \seight\ at 1.4~GHz using data from our JVLA
  program (see \S\ref{sec:radioobs}) and marking the `filament' and
  `swirl' referred to in this work.  An unassociated radio source lies
  to the NW.}
\label{fig:radio}
\end{figure}

Most of the best X-ray-studied cases are of radio galaxies of
relatively low radio power hosted by the brightest central galaxy in
the cluster or group environment, and less attention has been paid to
cases where there are multiple radio galaxies in relatively close
proximity. The radio galaxies hosted by \source\ and \seight\ are a
remarkable pair \citep{ekers} which reside in the central regions of
the cluster \cluster\ at $z=0.0381$ \citep{mazure} and have escaped
attention in recent years.  They were mapped at high resolution in the
radio with the VLA by \citet{cameron}, and an image showing the
salient features of Figure~\ref{fig:radio} appears in \citet*{bcg} in
the context of rotation-measure modeling of \seight.  The radio source
hosted by \source\ has asymmetric bent jets.  Lower-resolution radio
data show a very long bent extension on the jet side --- one of the
`tendrils' to which we refer later.  On the counterjet side there is
extreme looping, making a feature we refer to as the `swirl'.  The
radio source hosted by the eastern nucleus of the dumbbell galaxy
\seight\ is a classical double, of \citet{fr} type II (FR II)
morphology but unusual extended structure.  We refer to the bright
western extension from the southern lobe as the `filament'.  At lower
radio resolution, two long extensions (tendrils) are seen to the W and
NW.

While the complicated radio structures suggest interaction with the
cluster atmosphere, little has been published on \cluster's X-ray
properties.  It is detected in the \rosat\ All-Sky Survey (RASS) and
is one of the 447 members of the REFLEX cluster catalog
\citep{boehr04}.  However, it lies amongst the 12 per cent least
luminous REFLEX clusters, with a cataloged 0.1--2.4-keV luminosity of
only $1.8 \times 10^{43}$ ergs s$^{-1}$, and has not appeared in
samples for deeper study.  The \rosat\ High Resolution Imager pointed
at the field for 16.8~ks, and although investigation of the archival
data shows a clear detection of the centers of both galaxies, the data
are insufficiently sensitive for investigation of cluster
substructure.  In this paper we present new sensitive
\chandra\ observations of the system, together with new radio data
obtained with the Jansky Very Large Array (JVLA). \S\ref{sec:obs}
describes the new observations and data processing.
\S\ref{sec:cluster} and \S\ref{sec:galaxies} describe the cluster and
galaxy/group X-ray features, respectively.  In \S\ref{sec:rmorphs} we
highlight the fact that the two radio galaxies lie within the range of
power that dominates jet-mediated feedback in the Universe as a whole.
After examining available galaxy velocity data in
\S\ref{sec:velocities}, with reference to the distribution of cluster
gas and possible evidence for a merger, we discuss in
\S\ref{sec:discussion} relationships between the radio and X-ray
structures and likely underlying causes. \S\ref{sec:summary}
summarizes our results.  We adopt a luminosity distance for
\cluster\ of $D_{\rm L} = 168$ Mpc (appropriate for a Hubble constant
of 70 km s$^{-1}$ Mpc$^{-1}$), and 1 arcmin is equivalent to 45.3 kpc
at the source.  Flux densities and spectral indices are related in the
sense $S_\nu \propto \nu^{-\alpha}$.

\section{Observations \& analysis}
\label{sec:obs}

\subsection{X-ray}
\label{sec:xobs}

We made a 75~ks observation of the system in full-window and VFAINT
data mode with the Advanced CCD Imaging Spectrometer (ACIS) on board
\chandra\ on 2010 September 11 (OBSID 12241). \source\ was positioned
close to the nominal aimpoint of the front-illuminated I3 chip. The
other three CCDs of ACIS-I and the S2 chip of ACIS-S were also on
during the observations, giving a frame time of 3.14~s.  Details of
the instrument and its modes of operation can be found in the
\chandra\ Proposers' Observatory
Guide\footnote{http://cxc.harvard.edu/proposer}.  Results presented
here use {\sc ciao v4.5} and the {\sc caldb v4.5.6} calibration
database.  We re-calibrated the data to take advantage of the
sub-pixel event reposition routine ({\sc edser}), following the
software threads from the \chandra\ X-ray Center
(CXC)\footnote{http://cxc.harvard.edu/ciao}, to make new level~2
events files.  Only events with grades 0,2,3,4,6 were retained. After
screening to exclude intervals of high background at a threshold
appropriate for use of the blank-sky background files, the calibrated
dataset has an exposure time of 71.553~ks.

Since the cluster fills a large part of the detector array, background
was measured from blank-sky fields following procedures described in
the CXC software threads.  After cleaning the background data using
the same criteria as for the source data, and reprojecting to the same
coordinate system, a small normalization correction was applied (2\%)
so that the count rates matched in the $9.5-12$-keV energy band where
particle background dominates.

The {\sc ciao wavdetect} task was used to find point sources with a
threshold set to give 1 spurious source per field. Their regions were
subsequently masked from the data for the analysis of extended
structure.  All spectral fits are performed in {\sc xspec} on binned
spectra using $\chi^2$ statistics over the energy range $0.4 -
7$~keV.  The models include Galactic absorption of $N_{\rm H} =
5.27 \times 10^{20}$~cm$^{-2}$ \citep{dlock90}.  Parameter
uncertainties are 90\% confidence for 1 interesting parameter unless
otherwise stated.

\subsection{Radio}
\label{sec:radioobs}

We made sensitive, high-resolution, observations of the field
containing \source\ and \seight\ using the JVLA in its A configuration
at L (1.4~GHz) and C (5~GHz) bands (Table~\ref{tab:vlatab}).  At the
time of the observations, only part of the full bandwidth of the JVLA
correlator was available.  The data were calibrated and flagged for
extensive interference before being passed through the normal clean
and gain self-calibration cycles in {\sc casa}.  For more diffuse
structures we also downloaded archival L-band data taken in C
configuration with the VLA, and mapped them using standard procedures
in {\sc AIPS}.  Details of the heritage of the data sets and
properties of the resulting maps are given in Table~\ref{tab:vlatab}.
For L to C-band spectral-index measurements we made a version of the
C-band map with the same restoring beam as the L-band JVLA map.


\begin{deluxetable*}{lllllll}
\tablewidth{0pt}
\tablecaption{VLA/JVLA Radio Data}
\tablehead{
\colhead{Program} & \colhead{Observation Date} & \colhead{Frequency
  (GHz)} & \colhead{Configuration} & \colhead{Restoring Beam
  (arcsec)$^2$} & Noise (mJy beam$^{-1}$) &\colhead{Refs\tablenotemark{a}}
}
\startdata
AB1389 & 2011 Jun 27 & 1.39 & JVLA-A & $2.16 \times 1.01$ & 0.066 & 1\\
AB1389 & 2011 Jun 27 & 4.96 & JVLA-A & $0.86 \times 0.51$ & 0.056 & 1\\
AC105 & 1984 Mar 31 & 1.525 & VLA-C & $13.5 \times 13.5$ & 0.16 & 2\\
\enddata
\tablenotetext{a}{Publications also using data from these programs: 1 =
  Birkinshaw \& Worrall, in preparation. 2 = \citet{bcg}}
\label{tab:vlatab}
\end{deluxetable*}


\addtocounter{footnote}{1}
\let\thempfootnote\thefootnote

\begin{figure}
\centering
\includegraphics[width=1.0\columnwidth,clip=true]{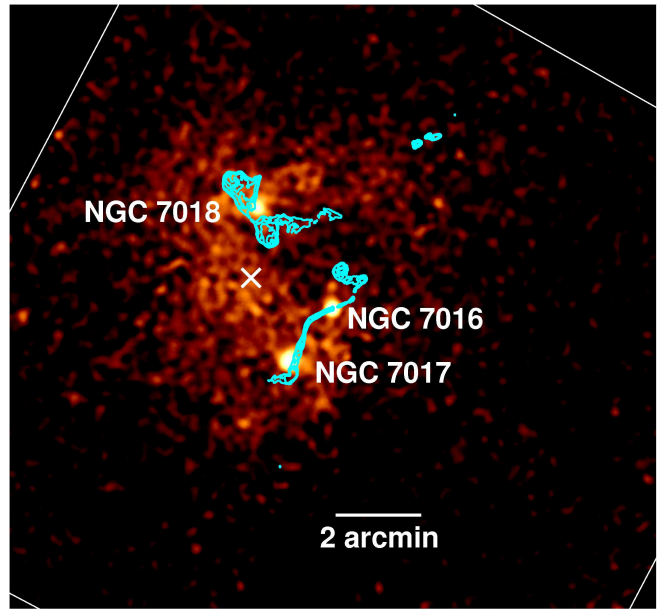}
\caption[]{0.3-5-keV exposure-corrected \chandra\ image in native
  0.492-arcsec pixels and smoothed in DS9\footnotemark ~with a 2D
  Gaussian of radius 20 pixels after removal of point sources.  Blue
  contours are at 0.5,1,2,4,8,16,32 mJy beam$^{-1}$ from a map of our
  1.4-GHz JVLA data made with a restoring beam of $2.16 \times 1.01$
  arcsec.  The outer lines mark the perimeter of the ACIS-I chip
  array.  The white cross is at $21^{\rm h} 07^{\rm m} 24^{\rm
    s}\llap{.}7, -25^\circ 27' 25''$, the position adopted
  as the center of the cluster gas.}
\label{fig:large}
\end{figure}

\footnotetext{hea-www.harvard.edu/RD/ds9/.
The standard deviation, $\sigma$, of the Gaussian is of size radius/2
pixels, and the smoothing kernel is truncated at $2\sigma$.
}


\section{X-rays from the cluster gas}
\label{sec:cluster}

Figure~\ref{fig:large} is a smoothed, exposure-corrected 0.3-5~keV
image of the \chandra\ data, after removal of point sources but not
the atmospheres of the three labeled NGC galaxies.  1.4-GHz radio
contours are overlayed.  The gaseous atmosphere of \sseven\ is seen in
projection on the S jet of \source.

The extended gas distribution very obviously deviates from spherical
symmetry.  Interestingly, the filament and swirl
(Fig.~\ref{fig:radio}) both correspond to regions where the X-ray
emission is less prominent.

\subsection{Integrated properties}
\label{sec:integrated}

\begin{figure}
\centering
\includegraphics[width=0.49\columnwidth,clip=true]{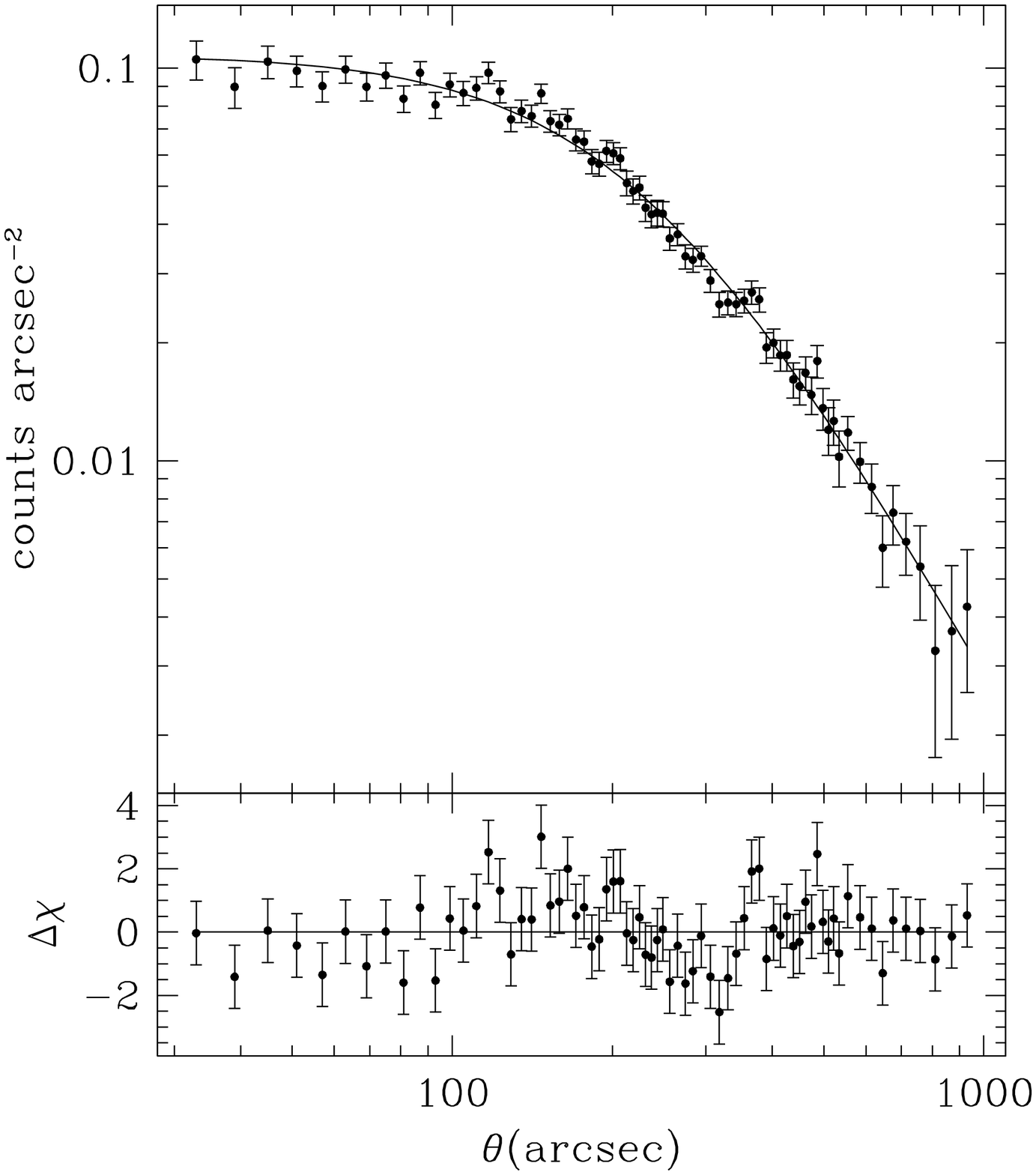}
\includegraphics[width=0.49\columnwidth,clip=true]{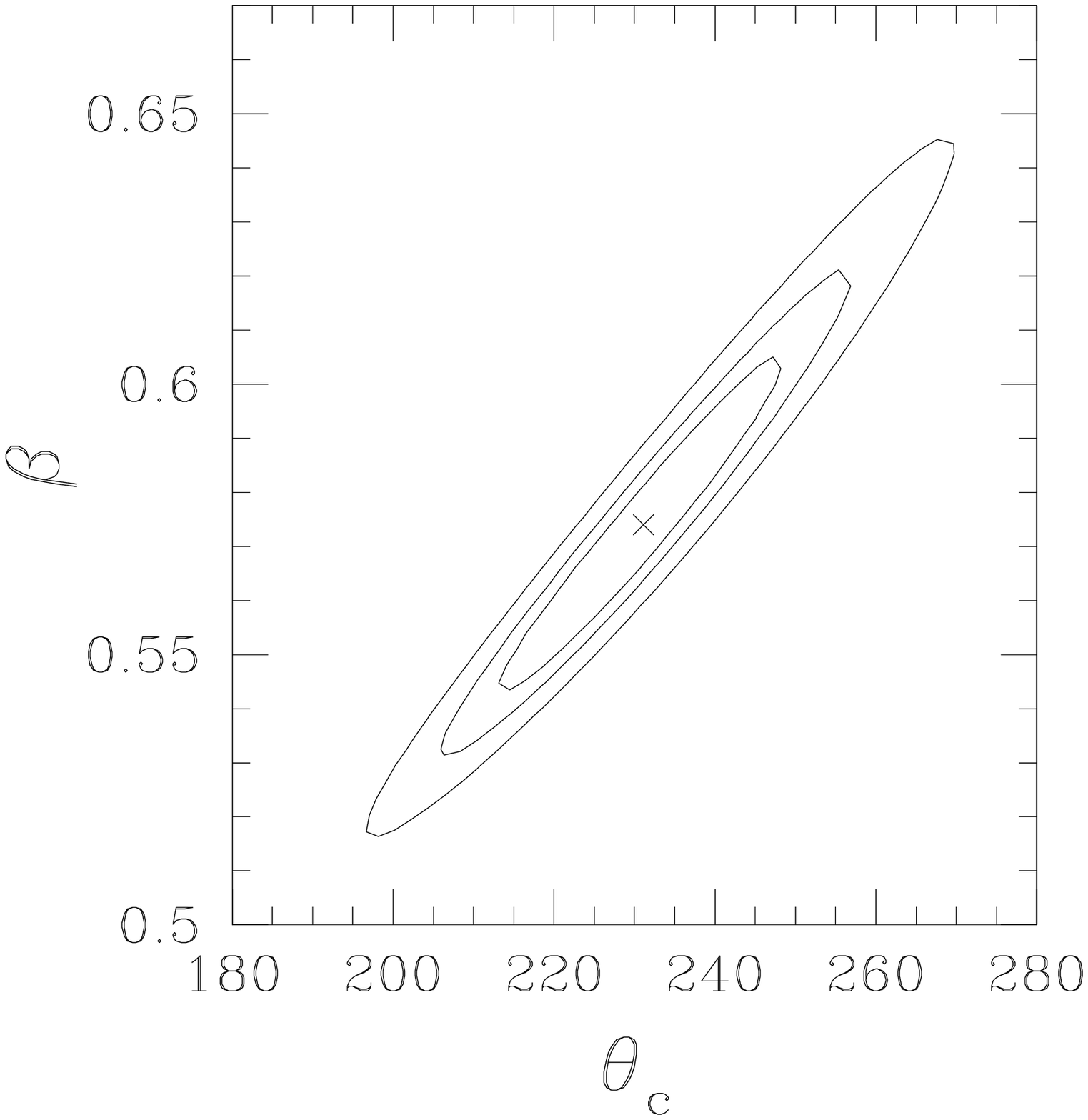}
\caption[]{{\bf Left:} Background-subtracted exposure-corrected radial
  profile fitted to a $\beta$-model profile with the residuals (shown
  as a contribution to $\chi$) in the lower panel.  The best fit is
  for a core radius of $\theta_{\rm c} =231.3$ arcsec and $\beta
  =0.574$ ($\chi^2/$dof = 84/69).  {\bf Right:} Uncertainty contours
  ($1\sigma$, 90\% and 99\% for 2 interesting parameters) of
  $\theta_{\rm c}$ and $\beta$ for the radial-profile fit.}
\label{fig:profile}
\end{figure}

Despite the lack of spherical symmetry, we have characterized the
overall extent of the cluster atmosphere by fitting a $\beta$-model
profile to a background-subtracted exposure-corrected radial profile
centered on the position $21^{\rm h} 07^{\rm m} 24^{\rm s}\llap{.}7,
-25^\circ 27' 25''$ (the approximate centroid of the diffuse
emission contained within a circle of radius 227 arcsec) shown as a
cross in Figure~\ref{fig:large}.  The result is shown in
Figure~\ref{fig:profile}.  We note the core radius of 220 arcsec
inferred by \citet{bcg} based on earlier, less sensitive, \rosat\ data
falls within the $1\sigma$ error bound for $\beta \approx 0.56$.

\begin{figure}
\centering
\includegraphics[width=0.6\columnwidth,clip=true]{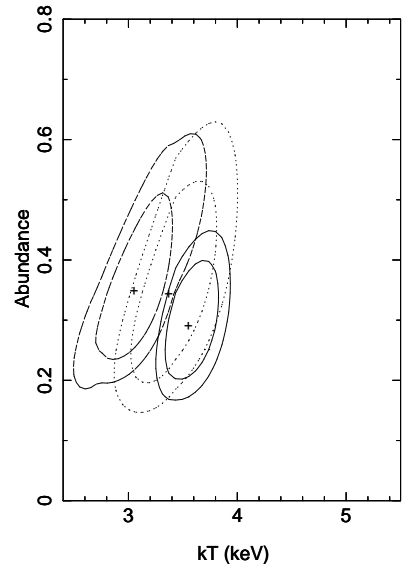}
\caption[]{ Spectral uncertainty contours ($1\sigma$ and 90\% for 2
  interesting parameters) of a fit of background-subtracted data (with
  point-sources and galaxy/group atmospheres removed) to a
  single-temperature thermal model.  From right to left: (a) solid
  contours: from a circle of radius 227 arcsec (best-fit $\chi^2$/dof
  = 352.3/348), (b) dotted contours: from an annulus of radii 227 and
  330 arcsec (best-fit $\chi^2/$dof = 315.6/319), and (c) dashed
  contours: from an annulus of radii 330 and 960 arcsec (best-fit
  $\chi^2/$dof = 485.9/448). }
\label{fig:clustertempabun}
\end{figure}

We have fitted the spectrum of the brightest extended emission in
Figure~\ref{fig:large} to a single-temperature thermal (APEC) model
absorbed by gas in the line of sight. Emission from the three
galaxy/group atmospheres has been excluded.  We find very similar
results for an on-source circle of radius 227 arcsec centered on
$21^{\rm h} 07^{\rm m} 24^{\rm s}\llap{.}7, -25^\circ 27' 25''$ as for
a polygon of similar area tracing better a contour of constant surface
brightness.  The spectral contours for the circular extraction region
and two outer annuli (radii 227 and 300 arcsec, and 300 and 960
arcsec) are shown in Figure~\ref{fig:clustertempabun}.  There is a
weak trend for the best-fit temperature to decrease with increasing
radius.  The 90\% confidence contour for the 227-arcsec-radius circle
and $1\sigma$ contour for the outer annulus do not touch, giving less
than 3 per cent probability of the spectral parameters being the same
in these regions.  The measurable difference is only in temperature,
with the abundances consistent with a constant value of roughly 0.3
times solar (i.e., $Z/Z_\odot$ = 0.3).

Beyond a radius of 330 arcsec some of the cluster lies off the ACIS
CCDs.  We have therefore estimated the total cluster luminosity using
a spectral fit to the emission within this circle, with point sources
and galaxy emission excluded, and have extrapolated to account for
missing flux based on the radial profile.  The spectrum gives an
acceptable fit to a model with $kT=3.50^{+0.16}_{-0.13}$~keV
($1\sigma$ error, see also Fig.~\ref{fig:clustertempabun}).  The
radial profile (Fig.~\ref{fig:profile}) finds that the counts within
330 arcsec should be multiplied by a factor of 2.39$\pm$ 0.09 to
account for emission out to $r_{500}$, corresponding to $\theta \approx 1200$
arcsec for the measured temperature based on \citet{vikhlinin}.  The
result is a bolometric luminosity for the cluster gas of (3.2 $\pm$
0.2) $\times 10^{43}$ erg s$^{-1}$ ($1\sigma$ error).  The luminosity
is too low to be a good match to the luminosity-temperature relations
of clusters --- see e.g., figure~5 of \citet{giles} --- or, in other
words, the temperature is too hot by $k\Delta T \approx 1.5$ keV for
the luminosity.  From the sample of \citet{maughan} at $0.1 <z < 1.3$,
the cluster with X-ray properties most closely resembling those of
\cluster\ is RXJ2247+0337 at $z=0.2$, with $kT =
2.7^{+0.7}_{-0.5}$~keV and a luminosity of $(4\pm1) \times 10^{43}$
erg s$^{-1}$ ($1\sigma$ errors).  That cluster is described as being
neither relaxed nor having a cool core --- an appropriate description
also of \cluster.

\cluster\ appears as RXC J2107.2-2526 in the REFLEX cluster catalog
\citep{boehr04}.  The 0.1-2.4~keV cluster luminosity of $1.8 \times
10^{43}$ erg s$^{-1}$ is based on 100 detected counts and used an
iterative method with an embedded temperature-luminosity relationship
which will have settled on a lower temperature than now measured. A
better-constrained value for the 0.1-2.4~keV luminosity to $r_{500}$
based on the \chandra\ data (calculated as above) is ($1.81
\pm 0.08$) $\times 10^{43}$ erg s$^{-1}$.

We find that the total gas mass out to $\theta=1200$ arcsec
is ($1.41 \pm 0.09$) $\times 10^{13}$ M$_\odot$, and
under the assumption of isothermal gas in hydrostatic equilibrium we
can use equations 30 and 32 of \cite{wbrev} to estimate a total mass
within this radius of ($1.94 \pm 0.15$) $\times
10^{14}$ M$_\odot$.  The relatively low gas-mass fraction is driven by
the unexpectedly high temperature that is measured (total mass is
proportional to temperature while the gas mass is largely independent
of it).  The central proton number density in the X-ray-emitting gas
is relatively low, at about 940 m$^{-3}$, which from figure~5 of
\citet{wbrev} we see corresponds to a long cooling time of about 48
billion years.

\subsection{Inner arm and cavity structures}
\label{sec:armhole}

\begin{figure}
\centering
\includegraphics[width=1.0\columnwidth,clip=true]{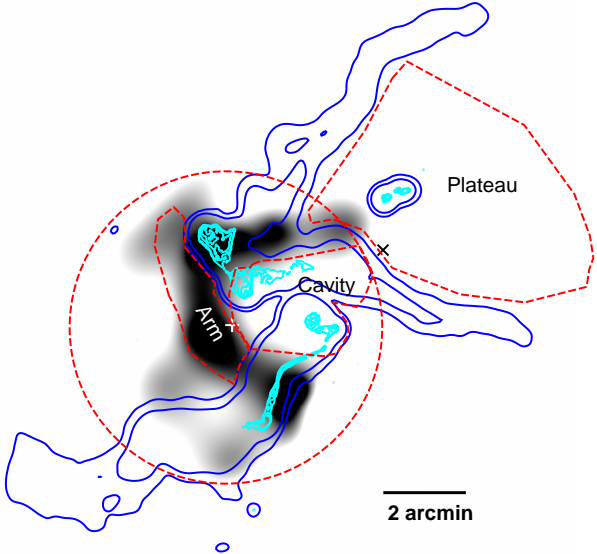}
\caption[]{Radio contours of Fig.~\ref{fig:large} are shown with outer
  contours at 0.5 and 5 mJy beam$^{-1}$ from a map with restoring beam
  of $13.5 \times 13.5$ arcsec made from the 1.5-GHz VLA data of
  program AC105 (solid line).  We refer to the three extended radio
  arms as `tendrils'.  The contours are overlayed on an adaptively
  smoothed grey-scale image showing the brightest region of cluster
  gas.  The black diagonal cross marks the center of the distribution
  of dominant cluster galaxies while the white cross is that for the
  small group whose center is closer to that of the gas distribution
  (\S~\ref{sec:velocities}). The dashed lines outline X-ray extraction
  regions: within the 227-arcsec-radius circle the arm region avoids
  extended radio emission while the cavity region does not.  }
\label{fig:armhole}
\end{figure}


\begin{figure}
\centering
\includegraphics[width=0.4\columnwidth,clip=true]{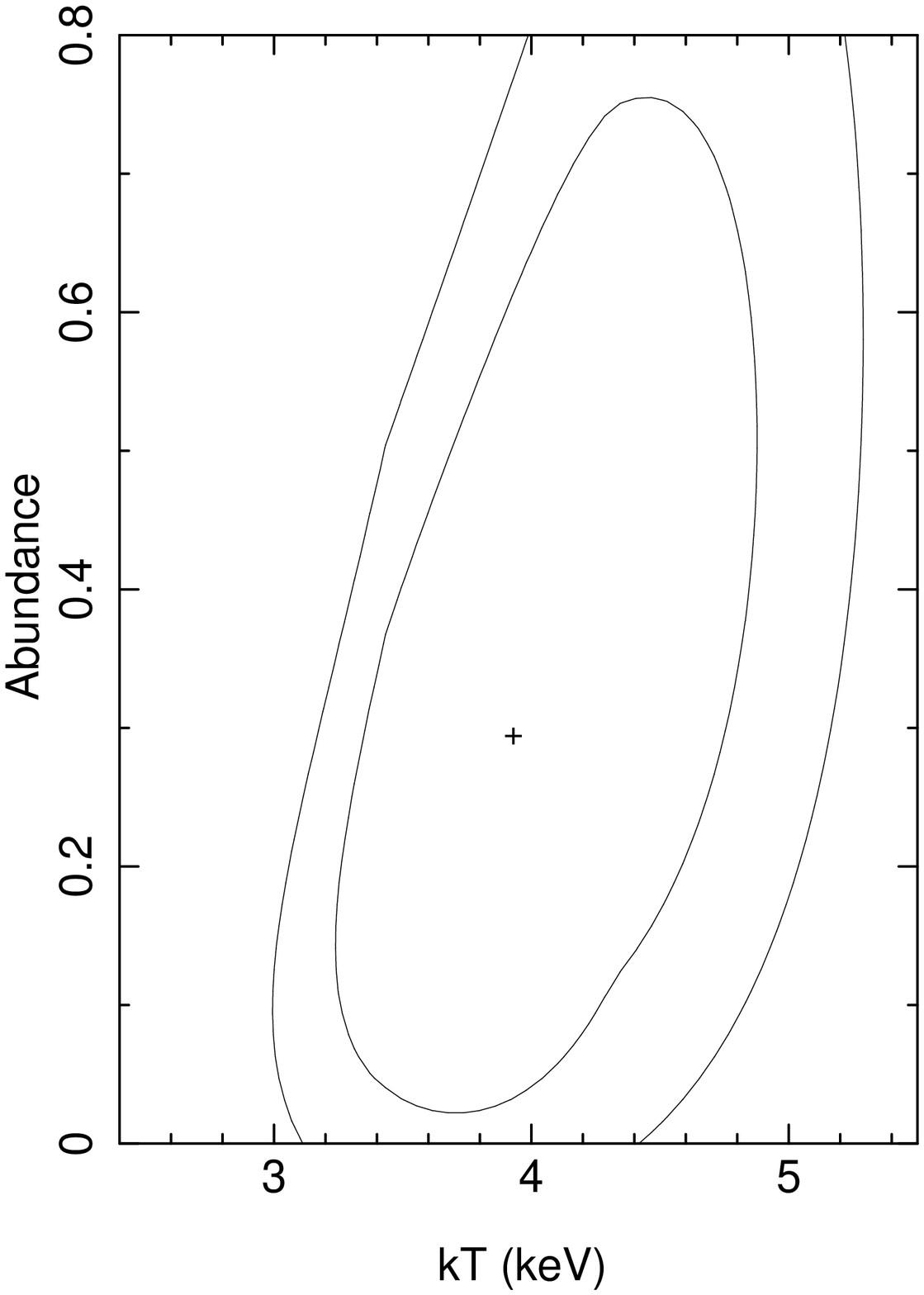}
\includegraphics[width=0.4\columnwidth,clip=true]{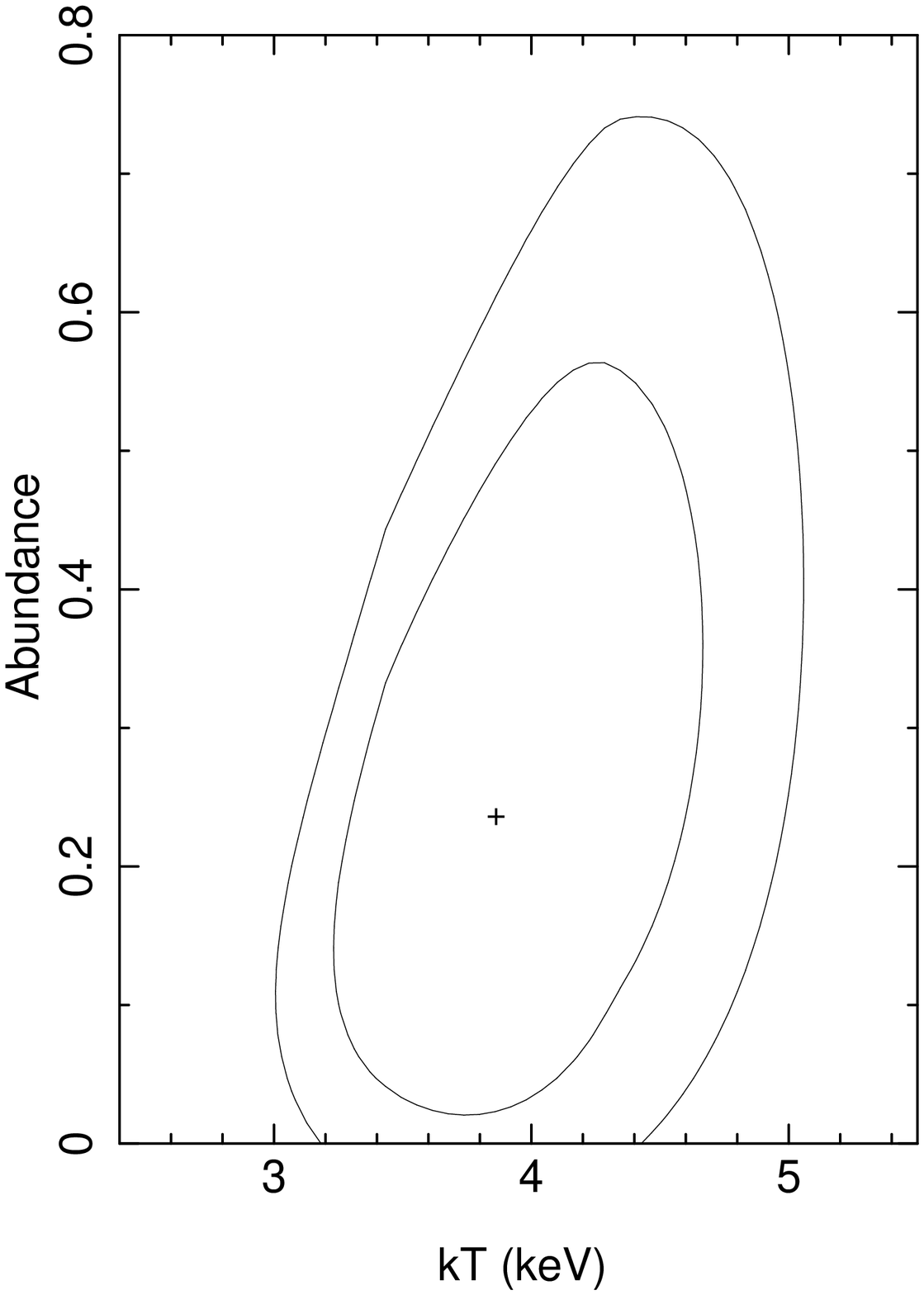}
\caption[]{As in 
Fig.~\ref{fig:clustertempabun} (including same axis scales)
for {\bf Left}: arm (best fit gives $\chi^2 = 71.1$ for 73
dof) and
{\bf Right}: cavity (best fit gives $\chi^2 = 103.5$ for 103
dof) 
shown in Fig.~\ref{fig:armhole}
}
\label{fig:armholetempabun}
\end{figure}


The gas morphology indicates a cavity encompassing the filament of
\seight\ and the swirl of \source, and so temperature structure in the
inner regions might be expected.  However, here we find no obvious
statistically-significant temperature structure and no cool core.  In
Figure~\ref{fig:armhole} we mark regions selected based on X-ray
and radio morphology that lie within the
227-arcsec-radius circle and that we refer to as the `arm' and
`cavity', and in Figure~\ref{fig:armholetempabun} we show their
spectral contours for temperature and abundance\footnote{The
  statistics for the cavity (1910 net counts) are slightly
  better than for the arm  (1528 net counts) 
  despite lower surface brightness.  This is because the arm has
  reduced exposure through lying over chip gaps for part of the
  telescope wobble.}.  The axes of Figure~\ref{fig:armholetempabun}
are the same as in Figure~\ref{fig:clustertempabun}.


 \begin{figure}
\centering
\includegraphics[width=0.7\columnwidth,clip=true]{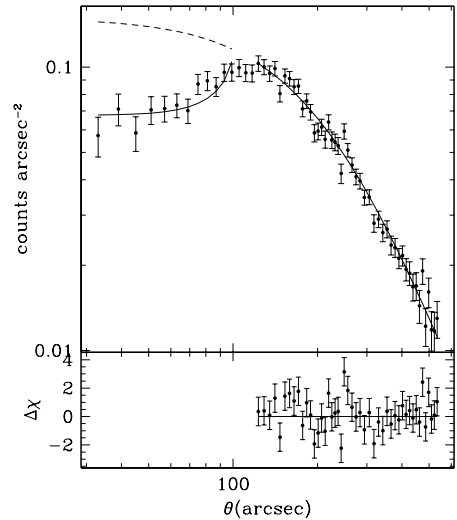}
\caption[]{ Background-subtracted exposure-corrected radial
    profile centered on the cavity and fitted to a $\beta$ model for
    $\theta > 120$ arcsec.  The dashed curve is the inner extension of
    the $\beta$ model.  The solid curve is the $\beta$ model from
    which has been subtracted the counts in the cavity which, since
    lying well within the core radius, is modeled as a
    zero-density sphere of radius 100 arcsec.}
\label{fig:cavityplotit}
\end{figure}


The lower average X-ray surface brightness towards the cavity with
respect to the arm (a factor of roughly 0.6) is difficult to
understand for gas of constant temperature.  
The apparent opening to the SW seen in Figure~\ref{fig:large}
  is very well aligned with the run of the chip gap, and so, despite
  exposure corrections having been applied, it is not possible to
  conclude that the region is unbounded in that direction.  We have
  therefore investigated what scale of spherical cavity is consistent
  with the data.  To do this we have extracted a radial profile
  centered roughly on the cavity, at $21^{\rm h} 07^{\rm m} 16^{\rm
    s}\llap{.}8, -25^\circ 26' 52''$.  As shown in Figure
  \ref{fig:cavityplotit}, the central data are clearly depressed
  relative to the extrapolation of the best-fit $\beta$ model fitted
  at angular radii larger than 120 arcsec (dashed line).

Since the
  angular scale of the counts deficit lies within the core radius, we
  have tested the simple model of a central evacuated spherical
  region by subtracting the profile of counts that would otherwise lie
  there (normalized to the central volume density of the $\beta$
  model) from the profile of the $\beta$ model.  Figure
  \ref{fig:cavityplotit} shows that such a central spherical
  evacuated region is consistent with observations if of angular
  radius roughly 100 arcsec.

It is not obvious how to construct a static cavity of this type in a
gas of constant pressure.  It would either need to be a dynamical
effect, or there needs to be a source of additional pressure in the
cavity.  It is likely that extra pressure is provided by
radio-emitting relativistic particles (note from
Fig.~\ref{fig:armhole} that radio emission covers the cavity but not
the arm), in which case a power-law X-ray component from
inverse-Compton scattering (discussed in \S\ref{sec:discuss-hole})
might be detectable.  To test how well a power law can be accommodated
within the region where the radio emission is brightest, we have
defined a further region, extending that covered by the cavity and
guided by the radio contours.  A spectral fit to either a
single-component thermal or power-law model is acceptable, at
$\chi^2/$dof = 105.1/94 and 106.7/95, respectively.  When the two
models are combined and the power-law slope is fixed to $\alpha =
0.9$, we find $\chi^2/$dof = 97.8/93 and the flux contribution of the
power law is $61^{+34}_{-37}$ per cent (90\% confidence).  While we
would not claim this as strong evidence for a power-law component, it
is consistent with a contribution of inverse Compton emission, as
discussed further in \S\ref{sec:discuss-hole}.

\subsection{The X-ray `plateau'}
\label{sec:plateau}

\begin{figure}
\centering
\includegraphics[width=0.8\columnwidth,clip=true]{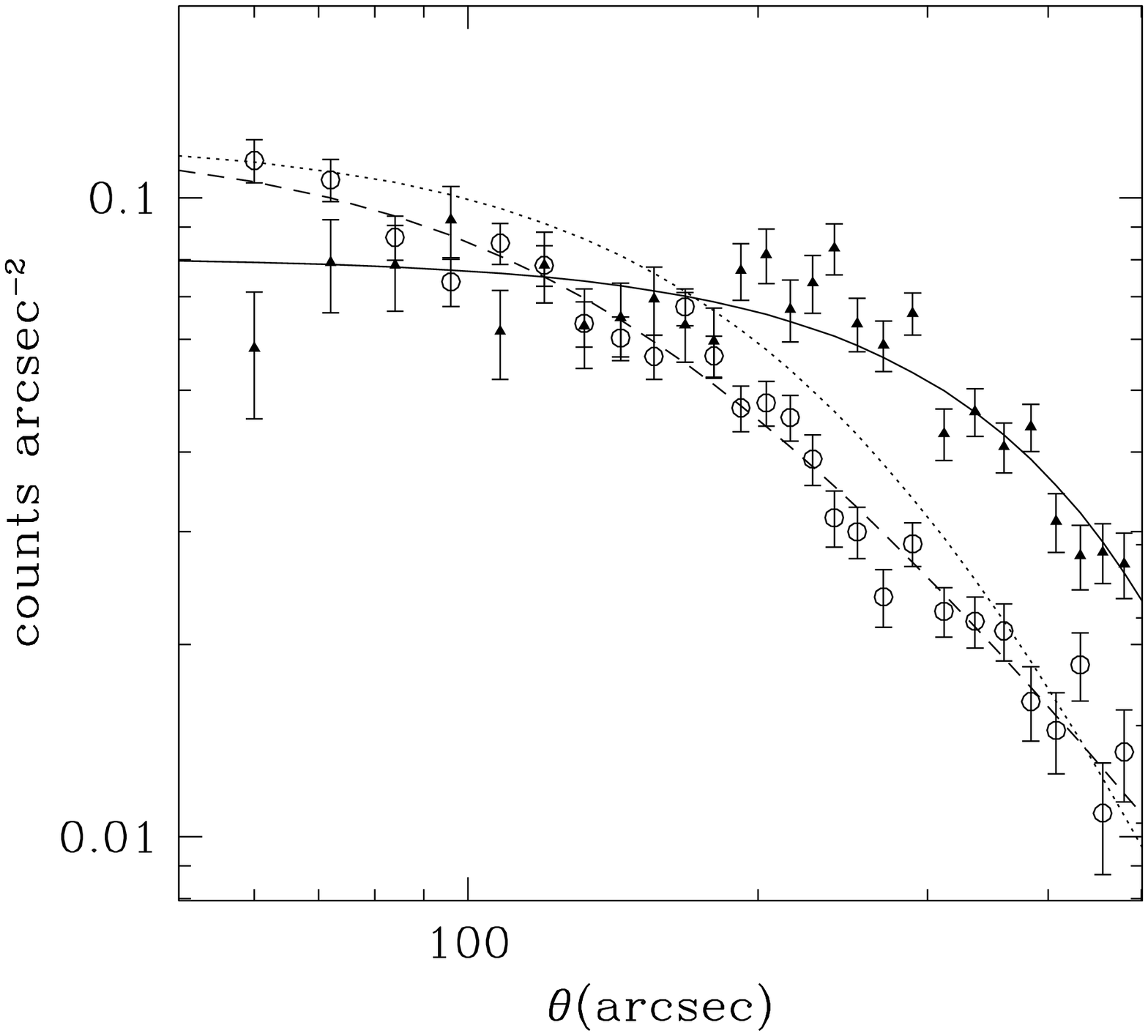}
\includegraphics[width=0.6\columnwidth,clip=true]{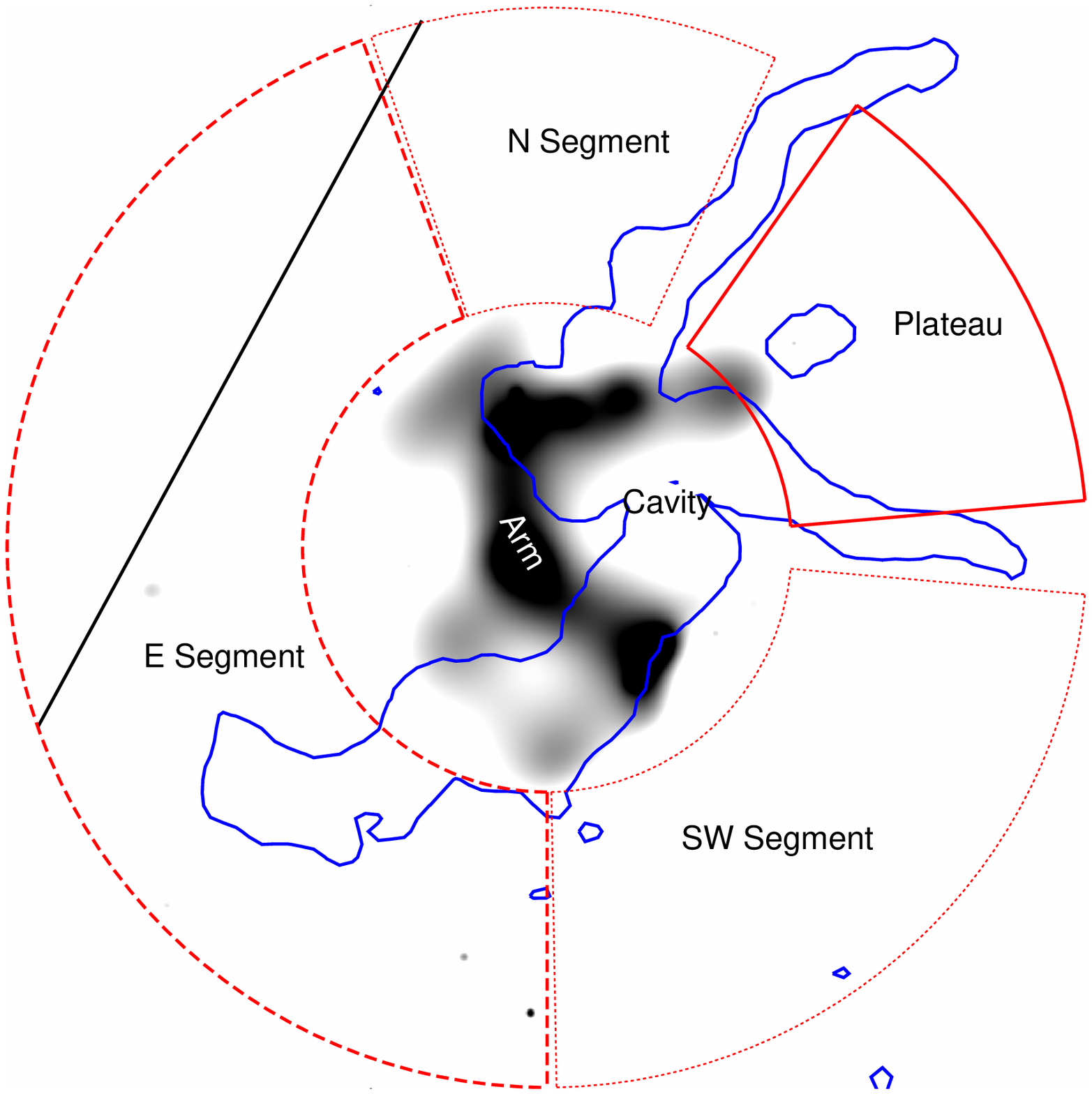}
\caption[]{{\bf Top:} Radial profile, as in the left panel of
  Fig.~\ref{fig:profile}, plotted separately for the subset of counts
  in pie slices of position angles 275--325 degrees (filled triangles,
  solid curve; plateau region) and 20--180 degrees (open circles,
  dashed curve; E segment).  The curves have the shapes of $\beta$
  models, but are only present to guide the eye. The dotted curve is
  for the combined N and SW segments (points suppressed). {\bf
    Bottom:} Adapted from Fig.~\ref{fig:armhole} to show the locations
  of the pie slices. The E segment is truncated by the edge of the
  detector (solid diagonal line).  }
\label{fig:profilesEW}
\end{figure}

The residuals in the radial-profile fit (Fig.~\ref{fig:profile}) show
structure at radii between 100 and 400 arcsec.  Profiles of different
pie slices were made, from which it became apparent that the main
feature was an X-ray excess that lies between \seight's tendrils of
1.5-GHz radio emission (Fig.~\ref{fig:armhole}) outside the gas
cavity.  We refer to this gas feature as the `plateau' and it is
marked in Figure~\ref{fig:armhole}.  Radial profiles for specific pie
slices (Figure~\ref{fig:profilesEW}) show excess counts in the profile
corresponding to the position angles of the plateau.

\begin{figure}
\centering
\includegraphics[width=0.6\columnwidth,clip=true]{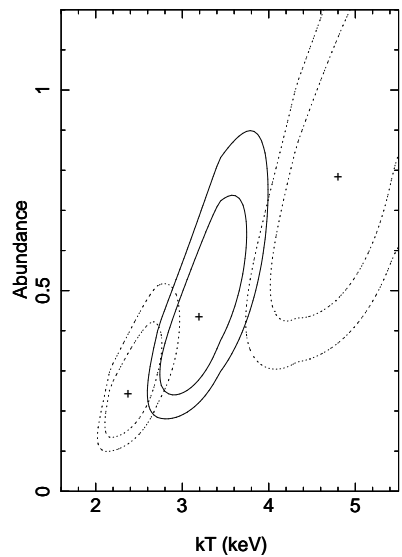}
\caption[]{ As in Fig.~\ref{fig:clustertempabun} for regions in lower
  panel of Fig.~\ref{fig:armholetempabun}.  Left to right: (a) E
  segment, dashed contours (3894 net counts, best-fit $\chi^2$/dof = 182.0/168), (b)
  plateau, solid contours (3225 net counts, best-fit $\chi^2/$dof = 181.3/204), and (c)
  combined N and SW segments, dotted contours (5152 net
    counts, best-fit $\chi^2/$dof =
  228.2/204). }
\label{fig:segtempabun}
\end{figure}

After identifying the plateau as a distinct morphological structure,
we investigated the spectral properties of the gas on these larger
scales in more detail.  While the gas in the plateau fits a
temperature similar to the average for the cluster, gas to the east at
radii beyond 227 arcsec was found to be cooler, and that to both sides
between the plateau and E segment was found to be hotter.  In
particular, the temperature of gas in the E segment
(Fig.~\ref{fig:profilesEW}) is cooler than that from the combined N
and SW segments at high significance (Fig.~\ref{fig:segtempabun}).

\begin{figure*}
\centering
\includegraphics[width=0.9\columnwidth,clip=true]{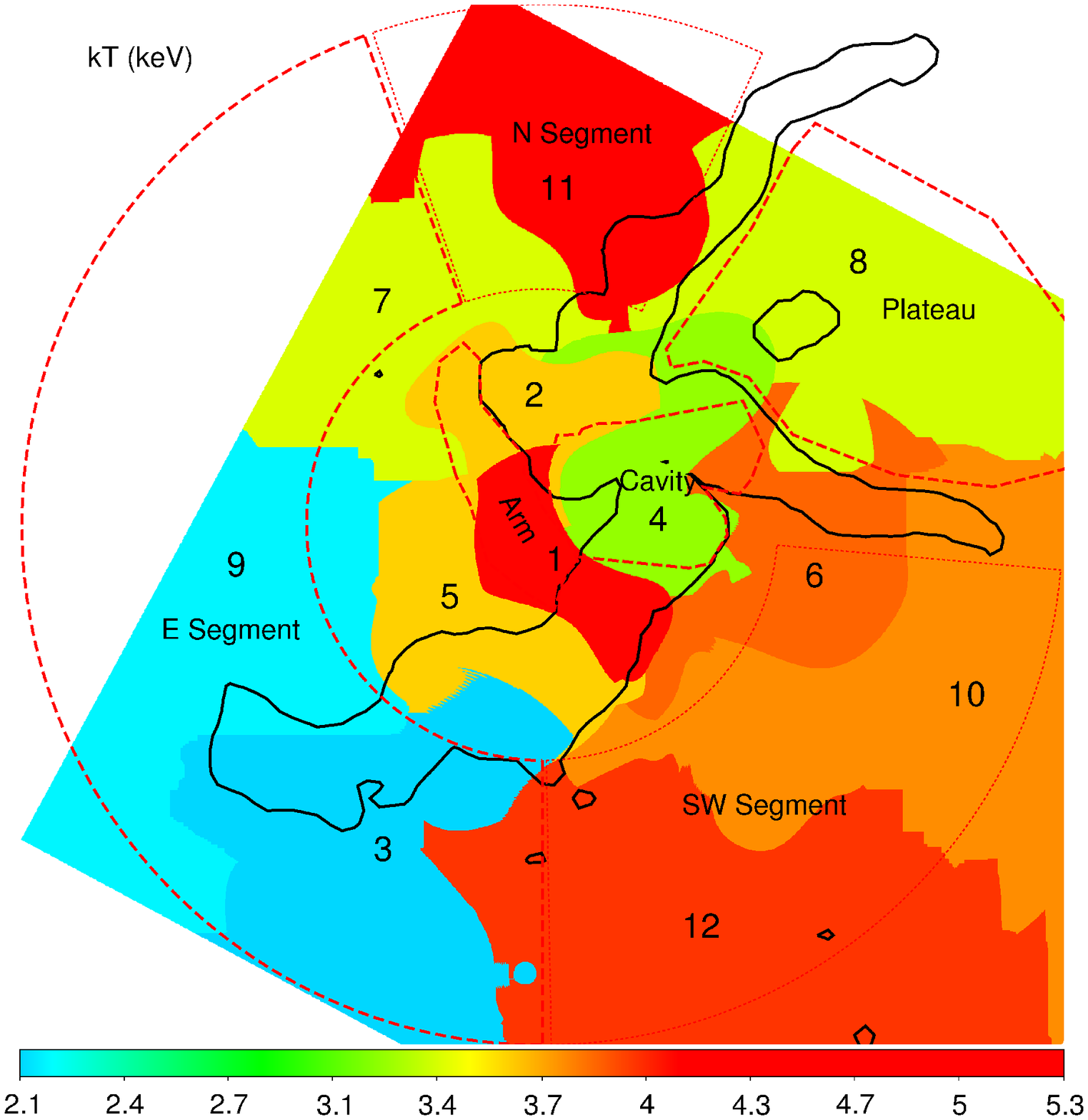}
\includegraphics[width=0.9\columnwidth,clip=true]{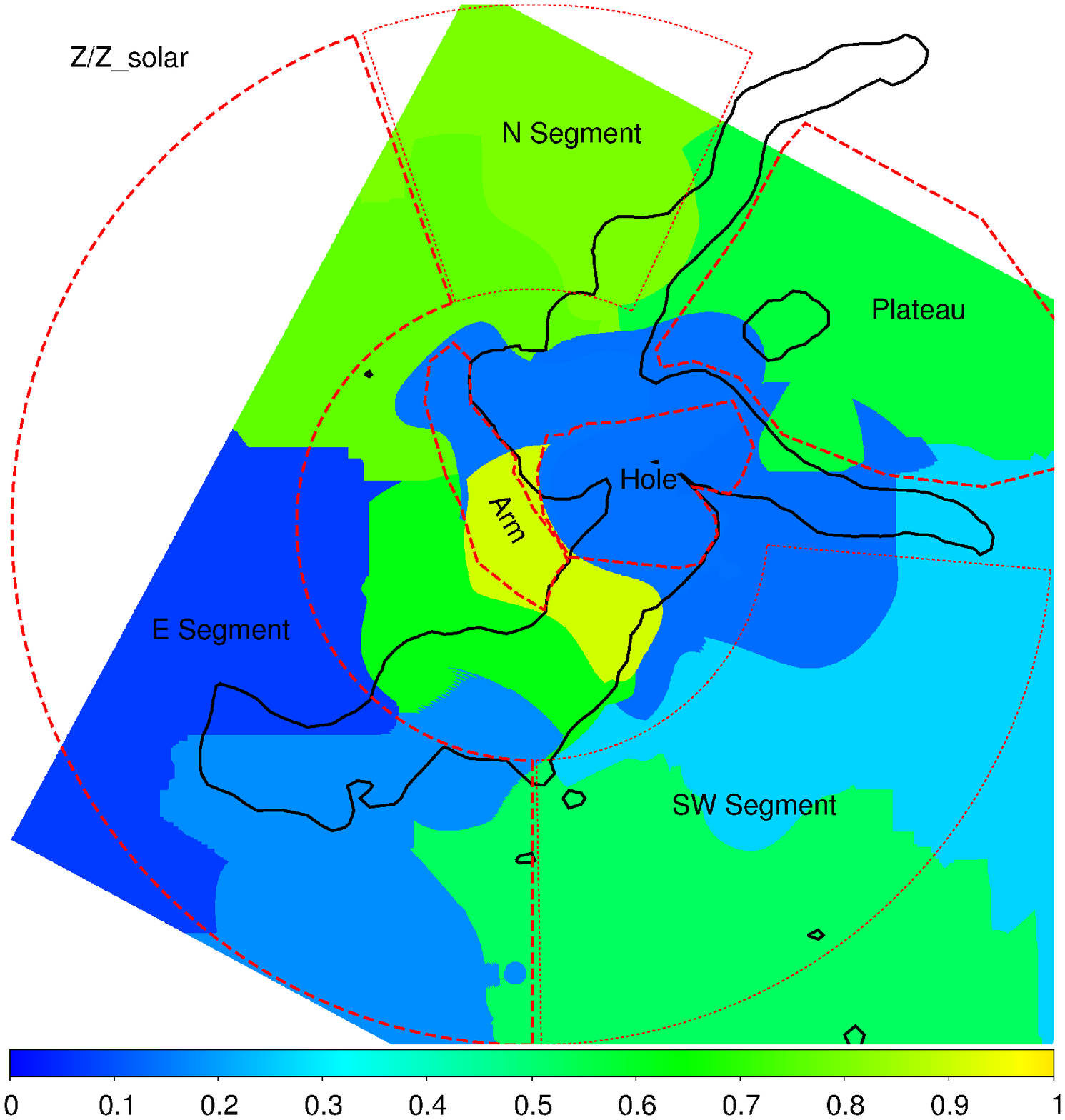}
\caption[]{Maps of temperature ({\bf left}) and metallicity relative
  to Solar ({\bf right}) for a single-component thermal model.}
\label{fig:contbin}
\end{figure*}

To probe further the locations of temperature changes, we have used
the {\sc  contbin}\footnote{http://www-xray.ast.cam.ac.uk/papers/contbin/}
software of \citet{sanders} to define spectral regions from our
adaptively-smoothed image, and map {\sc xspec} model-fitting results
onto the image.  
We used a constant signal to noise (S/N) based on our
  adaptively smoothed exposure-corrected (but not background
  subtracted) map, which provided regions for spectral fitting with a
  S/N ranging between 21 and 40.  The S/N was on purpose chosen to be
  slightly lower than that in the regions of Figure
  \ref{fig:segtempabun}, in order to investigate to what extent the
  sector boundaries of Figures \ref{fig:profilesEW} and
  \ref{fig:segtempabun} were supported by a less subjective
   examination of the data.
The fitted spectral model is a single-temperature thermal,
and we show results for $kT$ and metallicity in
Figure~\ref{fig:contbin}.

Spectral results are listed by region in Table~2.  The fits to
  the individual regions are acceptable.  When the outer regions of
most extreme temperature (3,6, 9, 11) are combined, we find an
unacceptable fit to a single-temperature model, as expected from
Figure~\ref{fig:segtempabun} which shows that gas in the E separates
in temperature from that in the N and SW.  Our main conclusions are
that gas to the E is cool ($kT$ about 2.1 keV) and of
metallicity less than about 0.36. Gas within a radius of 227 arcsec
and in the plateau is of intermediate temperature ($kT$ about
3.6 keV) and average cluster metallicity, except in a region
corresponding to the brightest radio emission from \sseven\ and
\seight, where a possibly depressed abundance might be due to dilution
with non-thermal power-law emission (\S\ref{sec:armhole} and
\ref{sec:discuss-hole}).  Gas wedged between the plateau and the cool
gas to the E is significantly hotter ($kT$ about 5 keV) and generally
of normal metallicity.

\begin{deluxetable}{lllcl}
\tabletypesize{\footnotesize}
\tablewidth{0pt}
\tablecaption{Spectral results for regions marked in Figure~\ref{fig:contbin}}
\tablehead{
\colhead{Region} & \colhead{$kT$ (keV)\tablenotemark{a}} & 
\colhead{$Z/Z_\odot$\tablenotemark{a}} & \colhead{$\chi^2/$dof} &
\colhead{Probability \tablenotemark{b}}
}
\startdata
1 & $4.11^{+0.68}_{-0.58}$ & $0.91^{+0.62}_{-0.43}$&103.5/92 & 0.19\\
2 & $3.60^{+0.59}_{-0.49}$ & $0.14^{+0.21}_{-0.14}$&103.8/108 & 0.60\\
3 & $2.09^{+0.50}_{-0.37}$ & $0.18^{+0.20}_{-0.11}$&73.9/79 & 0.64\\
4 & $3.21^{+0.55}_{-0.45}$ & $0.14^{+0.19}_{-0.13}$&155.9/138 & 0.14\\
5 & $3.59^{+0.64}_{-0.54}$ & $0.62^{+0.50}_{-0.32}$&86.5/113 & 0.97\\
6 & $3.86^{+1.13}_{-0.75}$ & $0.13^{+0.37}_{-0.13}$&56.3/56 & 0.46\\
7 & $3.38^{+0.75}_{-0.53}$ & $0.77^{+0.68}_{-0.37}$&120.2/102 & 0.11\\
8 & $3.37^{+0.76}_{-0.52}$ & $0.57^{+0.48}_{-0.27}$&85.6/88 & 0.55\\
9 & $2.18^{+0.68}_{-0.45}$ & $0.07^{+0.20}_{-0.07}$&114.8/95 & 0.08\\
10 & $3.76^{+1.24}_{-0.78}$ & $0.26^{+0.53}_{-0.26}$&101.4/105 & 0.58\\
11 & $5.20^{+2.00}_{-1.80}$ & $0.79^{+1.55}_{-0.67}$&73.2/67 & 0.28\\
12 & $3.98^{+1.71}_{-1.08}$ & $0.52^{+0.82}_{-0.39}$&57.8/66 & 0.75\\
3+6+9+11 & $3.00^{+0.39}_{-0.37}$ & $0.18^{+0.14}_{-0.10}$ &
342.4/292 & 0.02\\
\enddata
\tablenotetext{a}{90\% errors for one interesting parameter.}
\tablenotetext{b}{Null hypothesis probability. Unacceptable 
when 3,6,9,11 are combined.}
\label{tab:regspec}
\end{deluxetable}


\section{X-rays from the galaxies/groups}
\label{sec:galaxies}

We have extracted spectra of the X-ray atmospheres surrounding the
three galaxies marked on Figure~\ref{fig:large} using local background
from a surrounding annulus rather than the blank-sky background, as
the galaxies and their group atmospheres are embedded within cluster
gas.  \source\ and \seight\ both lie on the I3 CCD of ACIS, and
\sseven\ is on I2.  We have used 2MASS J-band data accessed with {\sc
  skyview}\footnote{http://skyview.gsfc.nasa.gov/skyview.html} to
associate X-ray and radio features with their host galaxies.

\subsection{NGC 7016}
\label{sec:n7016}

\begin{figure}
\centering
\includegraphics[width=0.9\columnwidth,clip=true]{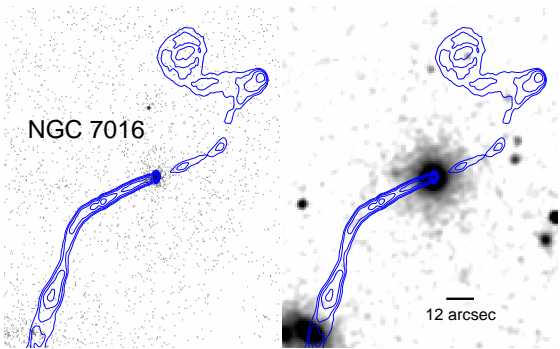}
\caption[]{ Radio contours of Figure~\ref{fig:large} for \source\ on
  {\bf Left} unsmoothed 0.3-5 keV X-ray image in native pixels and
  {\bf Right} 2MASS J-band image.  The host galaxy does not have a
  companion of similar brightness.  The swirl has not obviously been
  shaped by a feature seen in the X-ray or near IR.  }
\label{fig:n7016X-2MASS}
\end{figure}

\begin{figure}
\centering
\includegraphics[width=0.7\columnwidth,clip=true]{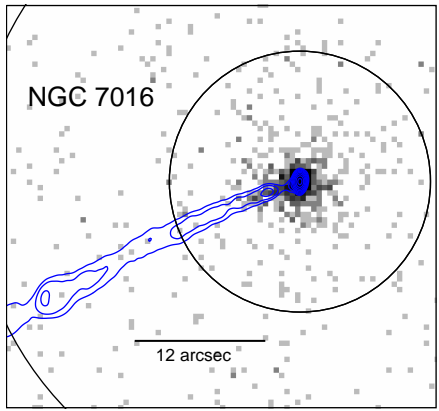}
\caption[]{Unsmoothed 0.3-5 keV X-ray image of \source\ in native
  0.492-arcsec pixels, showing that the nuclear emission is extended
  along the inner part of the SE radio jet.  Contours are at 0.15 mJy
  beam$^{-1}$, increasing by factors of 2 to 38.4 mJy beam$^{-1}$,
  from a map of our 5-GHz JVLA data made with a restoring beam of
  $0.86 \times 0.51$ arcsec.  The circle of radius 12 arcsec indicates
  the on-source spectral extraction region, with background taken from
  the partially shown annulus of radii 12 and 30 arcsec.  }
\label{fig:n7016}
\end{figure}

NGC\,7016 is a relatively isolated galaxy
(Fig.~\ref{fig:n7016X-2MASS}).  The 886 net counts from a
12-arcsec-radius source-centered circle (Fig.~\ref{fig:n7016}) give a
poor fit to a thermal model alone ($\chi^2/$dof = 79.9/30), but the
fit is acceptable when a power law (with no excess absorption) is
added to account for emission from the active nucleus.  The abundance
is poorly constrained and was fixed at 0.3 times solar in the
two-component fit.  Results are given in Table~\ref{tab:gspectab}. The
temperature is cooler than that of the surrounding cluster, as
expected for a galaxy atmosphere.

From a smaller circle of radius 1.5 arcsec, again centered on the
radio-core position of $21^{\rm h} 07^{\rm m} 16^{\rm s}\llap{.}28,
-25^\circ 28' 08''\llap{.}5$, the counts are more highly dominated
by the power-law component and spatially sharply peaked, but the fit
was still improved by a small contribution from thermal gas (from
$\chi^2/$dof = 26.9/17 to 10.7/15).  The power-law parameter values
were consistent with those from the larger extraction region,
demonstrating an active-nucleus origin of the non-thermal
emission. The power-law X-ray flux density (Table~\ref{tab:gspectab})
and 4.96-GHz core flux density of 58.3~mJy lie within the scatter of
the correlation of unabsorbed nuclear emission components of
\citet{evans}, arguing in favor of a common non-thermal origin.  An
extension of the X-ray emission in the direction of the radio jet
(Fig.~\ref{fig:n7016}) strongly suggests X-ray synchrotron emission
from an inner radio jet, found by \chandra\ to be a common feature of
FR\,I radio galaxies \citep*{worrall01}.

\subsection{NGC 7017}
\label{sec:n7017}

\begin{figure}
\centering
\includegraphics[width=0.9\columnwidth,clip=true]{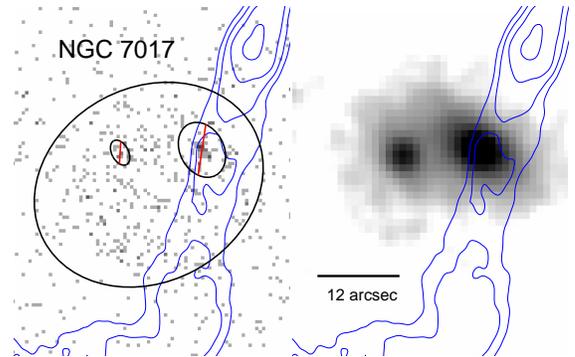}
\caption[]{ Radio contours of Figure~\ref{fig:large} shown (for
  reference only) projected on \sseven\ in {\bf Left} unsmoothed 0.3-5
  keV X-ray image in native pixels, with spectral extraction region
  shown and {\bf Right} 2MASS J-band image.  X-ray emission is seen
  predominantly from the envelope of the galaxy pair, with excess
  counts associated with the core of each.  }
\label{fig:n7017X-2MASS}
\end{figure}

There is a bright pair of galaxies associated with
\sseven\ (Fig.~\ref{fig:n7017X-2MASS}), although their velocity
difference of $2505\pm 36$ km s$^{-1}$ \citep{qr} is large.  The
centers of each were detected as X-ray point sources by {\sc
  wavdetect}.  We excluded the counts in circles of radii 3 and 1.2
arcsec around the centers of the western and eastern galaxy,
respectively, and 176 net counts were then detected from an ellipse of
semi-axis lengths of 17.4 and 14.2 arcsec shown in
Figure~\ref{fig:n7017X-2MASS}.  A good fit to a thermal model
(Table~\ref{tab:gspectab}) was found.

\subsection{NGC 7018}
\label{sec:n7018}

\begin{figure}
\centering
\includegraphics[width=0.9\columnwidth,clip=true]{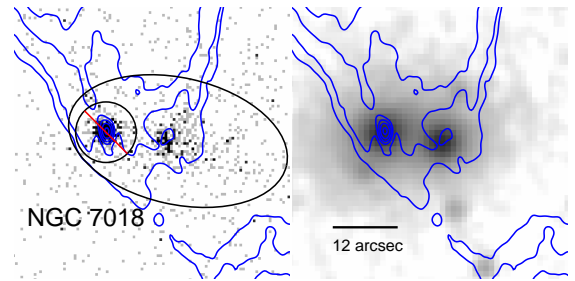}
\caption[]{ Radio contours of Figure~\ref{fig:large} for \seight\ on
  {\bf Left} unsmoothed 0.3-5 keV X-ray image in native pixels, with
  spectral extraction region shown and {\bf Right} 2MASS J-band image.
  The eastern galaxy of the pair hosts the core of the radio source
  together with predominantly point-like X-ray emission.  The western
  galaxy is associated with a weak radio excess and predominantly
  diffuse X-ray emission.  }
\label{fig:n7018}
\end{figure}

NGC\,7018 is also a galaxy pair.  The eastern member hosts the
radio-galaxy nucleus, and here point-like nuclear X-ray emission that
strongly outshines diffuse gas emission is seen
(Fig.~\ref{fig:n7018}).  709 net counts from a circle of radius 5.6
arcsec centered on $21^{\rm h} 07^{\rm m} 25^{\rm s}\llap{.}65,
-25^\circ 25' 43''\llap{.}3$ can be fitted well to a power-law model
with no excess absorption (Table~\ref{tab:gspectab}).  As for
\source\ the power-law X-ray flux density (Table~\ref{tab:gspectab})
and 4.96-GHz core flux density of 36.0~mJy lie within the scatter of
the correlation of unabsorbed nuclear emission components of
\citet{evans}, arguing in favor of a common non-thermal origin.

\citet{qr} find a velocity difference of only $117\pm30$ km s$^{-1}$
between the radio host galaxy and its western companion.  Diffuse
X-ray emission, associated largely with the companion galaxy, is seen
to the west of the radio nucleus.  385 net counts were extracted from
an ellipse of semi-axis lengths 20.6 and 11.7 arcsec, excluding the
circle described above and shown in Figure~\ref{fig:n7018}.  The data
give a good fit to a thermal model (Table~\ref{tab:gspectab}).  We
note that there appears to be a weak excess of radio emission
associated with the western galaxy.

\begin{figure}
\centering
\includegraphics[width=0.9\columnwidth,clip=true]{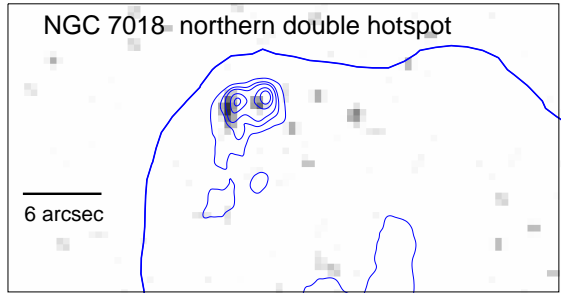}
\caption[]{ Outer 1.4-GHz radio contour of Figure~\ref{fig:large}
  outlining the northern edge of the northern lobe of \seight\ (bold
  line) together with contours from our higher-resolution 5-GHz map at
  0.3, 0.9, 1.5, 3, and 4.5 mJy beam$^{-1}$, showing the double
  hotspot.  X-rays are seen from both parts of the double hotspot: the
  image is 0.3-5-keV \chandra\ data in native pixels smoothed in DS9
  with a 2D Gaussian of radius 2 pixels.  The apparent misalignment of
  X-ray and radio emission may be due to inaccuracies in the 5-GHz
  radio mapping 3.5 arcmin off axis.  }
\label{fig:n7018nhots}
\end{figure}

The radio core of \seight\ is within its northern lobe.  While a broad
trunk of radio emission connects it to the southern lobe, the jet to
the north is narrow and more distinct, terminating in a double hotspot
(Fig.~\ref{fig:radio}).  There is excess X-ray emission to the north
of the nucleus associated with the tip of the inner jet
(Fig.~\ref{fig:n7018}; 14 counts detected in a circle of radius 2
arcsec where an average of 2.5 is expected based on local background).
The northern hotspot region is also detected in X-rays, with 18 counts
detected in a circle of radius 3.5 arcsec where 5 counts are expected
based on local background.  Hotspot counts appear to be associated
with each component of the double system, with a small misalignment
between X-ray and radio that is probably within the tolerance of
off-axis mapping uncertainties in the radio
(Fig.~\ref{fig:n7018nhots}).  The southern hotspot region is
undetected in X-rays.

\begin{deluxetable*}{llcccccc}
\tabletypesize{\footnotesize}
\tablewidth{0pt}
\tablecaption{Galaxy X-ray Spectral Results}
\tablehead{
\colhead{Region} & \colhead{$z$\tablenotemark{a}} & \colhead{$kT$ (keV)} &
\colhead{$Z/Z_\odot$} & \colhead{$N$ (cm$^{-5}$)\tablenotemark{b}} & 
\colhead{$\alpha_{\rm x}$} & \colhead{$S_{\rm 1~keV}$ (nJy)} &
\colhead{$\chi^2$/dof}
}
\startdata
\source\ & 0.03685 & $0.87^{+0.08}_{-0.06}$ & $0.3$ (fixed) & $5.1^{+1.2}_{-1.1} \times
10^{-5}$ & $0.72^{+0.38}_{-0.45}$ & $8.6^{+4.0}_{-3.5}$ & 29.9/29\\
\sseven\ & 0.03465 & $1.14^{+0.18}_{-0.21}$ & $0.3^{+1.2}_{-0.2}$ & $2.5^{+2.3}_{-1.8} \times
10^{-5}$ & - & - & 7.2/8\\
\seight-nucleus & 0.03842 & - & - & - & $0.75\pm 0.14$ & $15.8\pm 1.6$ & 26.9/25\\
\seight-gas & -  & $1.01^{+0.14}_{-0.10}$ & $0.14^{+0.11}_{-0.06}$ & $7.0^{+2.4}_{-2.0} \times
10^{-5}$ & - & - & 17.5/14\\
\enddata
\tablenotetext{a}{Redshift from NED of galaxy with the brightest X-ray
  emission (western component of \sseven\ and eastern component of \seight).}
\tablenotetext{b}{$10^{14} N = {(1+z)^2 \int n_{\rm e} n_{\rm p} dV
\over 4\pi D_{\rm L}^2}$, where $n_{\rm p}, n_{\rm e}$ are densities of
hydrogen nuclei and electrons, respectively, $V$ is volume and $D_{\rm
L}$ is luminosity distance.}
\label{tab:gspectab}
\end{deluxetable*}


\subsection{NW radio source}
\label{sec:nwrsource}

\begin{figure}
\centering
\includegraphics[width=0.9\columnwidth,clip=true]{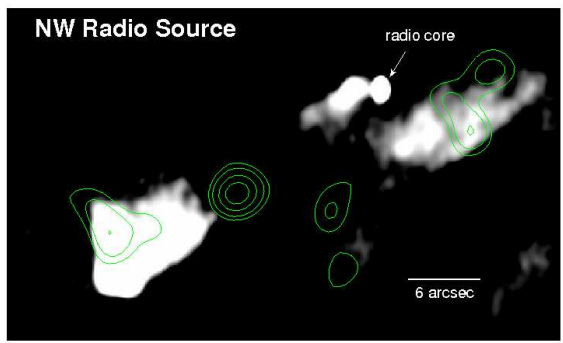}
\caption[]{ 4.96-GHz image of the distorted radio source to the NW of
  the field, with core at $21^{\rm h} 07^{\rm m} 06^{\rm s}\llap{.}11,
  -25^\circ 24' 04''\llap{.}4$.  The contours are a crude
  representation of groupings of X-ray counts, made from a lightly
  smoothed image.  The X-rays at the NW tip of the SE lobe represent a
  real source containing roughly 7 excess counts (0.3-5 keV), but it
  is most likely unrelated to the radio source.  }
\label{fig:nwradio}
\end{figure}

The small radio source in the NW of the field of
Figure~\ref{fig:radio} has a curious structure, as shown in
Figure~\ref{fig:nwradio}.  {\sc wavdetect} finds an X-ray source at
the NW tip of the SE lobe with about 7 excess counts (0.3-5 keV).
This seems to be chance coincidence, since it does not match the
location of the radio core or any other radio feature.  There is
possible excess X-ray emission associated with bright parts of the
lobes, but it is not possible to establish a level of significance
since X-ray counts are generally sparse in this part of the image.
There is no strong 2MASS association of the radio source.

\section{The radio-galaxy morphologies}
\label{sec:rmorphs}

The flux density of \seight\ at 1.525~GHz is roughly 6.2 Jy, of which
about 80 per cent lies in the classical-double structure and the rest
in its tendrils: a NW tail of the N lobe and the W extension of the
radio filament of the S lobe.  While data at lower frequencies have
been published, the large beam sizes confuse the structures, and total
flux densities combine emission from \source\ and \seight.  Adopting a
spectral index of $\alpha_{\rm r}=0.9$ we extrapolate to find a
178-MHz power for the double source \seight\ of $9 \times 10^{24}$ W
Hz$^{-1}$ sr$^{-1}$, placing it in the range where Fanaroff and Riley
types I and II (FR\,Is and FR\,IIs) overlap in luminosity \citep{fr}.
These are the sources with the correct range of power to dominate
jet-mediated feedback in the Universe as a whole, and so are of
particular importance \citep{wrev}.  The best-studied example is
PKS~B2152-699, where not only have the lobes evacuated cavities in the
X-ray-emitting gas, but also relatively strong (Mach number between 2
and 3) shocks are seen, and the kinetic and thermal energy of shocked
gas dominates over the cavity power \citep{wfos}.  In \seight\ the S
lobe may have helped to bore out the large X-ray cavity in which it
resides (Fig.~\ref{fig:armhole}, and see below), but there is no
evidence for shocked gas.  There are differences in radio morphology
that may explain why strong shocks are not detected around \seight.
In particular, \seight\ has bright terminal hotspots rather than the
lobe-embedded hotspots seen in PKS~B2152-699, perhaps suggesting that
strong shocks are detected more readily either during a phase of the
source's evolution or at preferred source orientations.  X-ray study
of more sources of these radio powers are needed to resolve the
issues.

\source, with a total flux density of 4.9 Jy at 1.525 GHz, is almost
as bright as \seight, although it exhibits no FR\,II morphology.  Just
over half of the flux density is from the region beyond the sharp
southern bend of the SE jet that is seen well in the high-resolution
radio data, and about 64 per cent of that is in the region of the
tendril that lies beyond the extent of the jet in our high-resolution
map.

\section{Galaxy velocities}
\label{sec:velocities}

The gas of \cluster\ is highly disturbed.  Merging can produce
irregularities, and so we have examined the available galaxy
velocities.  In order to search for sub-clumps we used velocities
listed in NED for galaxies within 30 arcmin of \source, noting that
the mean redshift of $0.03824 \pm 0.00024$ and velocity dispersion of
$573^{+43}_{-55}$ km s$^{-1}$ of 72 galaxies in the field, and likely
associated with the cluster, are in good agreement with the redshift,
and dispersion of 559 km s$^{-1}$, found by \citet{mazure}.  We
applied an algorithm that looks for objects close to one another in
three dimensions, ignoring velocity differences less than 800 km
s$^{-1}$.  The search scale for these associations begins at 1 kpc and
increases slowly to 1 Mpc, and pairs are combined into lumps on each
scale at their mean position and velocity.  The input parameters were
then adjusted to check for stability in the results.

\begin{figure}
\centering
\includegraphics[width=0.9\columnwidth,clip=true]{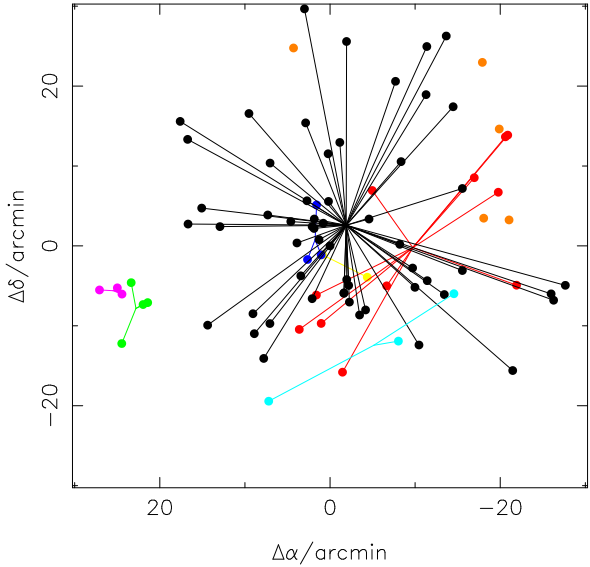}
\caption[]{ Tree analysis of galaxy associations based on position and
  velocity.  The plot is centered at the position of \source.  There
  are 56 galaxies associated with a single cluster grouping, and other
  groups are sparse. Galaxies unassociated with others are shown as
  dots with no lines.  }
\label{fig:tree}
\end{figure}

As shown in Figure~\ref{fig:tree}, the dominant structure is
  a cluster of 56 galaxies centered at $21^{\rm h} 07^{\rm m} 07^{\rm
    s}\llap{.}9, -25^\circ 25' 33''$, about 4 arcmin west of the
  nucleus of \seight, with mean velocity 11290 km s$^{-1}$ and
  velocity dispersion 350 km s$^{-1}$.  The cluster includes
  \source\ (11046 km s$^{-1}$) and both nuclei of \seight\ (11517 and
  11694 km s$^{-1}$), which are among the bright galaxies in the
  cluster, but none of which is clearly identifiable as a central
  brightest cluster galaxy.  Several groups can also be identified,
  including one of velocity 13030 km s$^{-1}$ associated with the
  eastern component of \sseven\ (12892 km s$^{-1}$) and two other
  galaxies.  This group, at roughly $21^{\rm h} 07^{\rm m} 24^{\rm s},
  -25^\circ 27' 22''$, is closer to the center of the gas distribution
  than that of the cluster galaxies.  Both centers are marked in
  Figure \ref{fig:armhole}.  The second largest grouping is of eleven
  galaxies centered about 8 arcmin to the WSW, at $21^{\rm h} 06^{\rm
    m} 33^{\rm s}, -25^\circ 28' 21''$ with velocity 12140 km
  s$^{-1}$.  The sparseness of velocity information for galaxies that
  might be in these distinct groups makes it difficult to interpret
  the neighborhoods of \source\ and \seight\ with certainty, but
  interactions of the group associated with the western component of
  \sseven\ and the main cluster also seems plausible.  This component
  of \sseven\ is the brightest galaxy in the field but, at a velocity
  of 10387 km s$^{-1}$, appears to be dynamically distinct from the
  main cluster, and may have an associated infalling group of galaxies
  without measured velocities.

The position of the main galaxy centroid explains the excess plateau
gas, although not why it is so well positioned between the radio
tendrils (Fig.~\ref{fig:contbin}).  The velocity data are suggestive
that a merger interaction between a group or groups with significant
velocity offsets from the main cluster is responsible
for the non-uniformities in gas distribution and temperature.

\section{Discussion}
\label{sec:discussion}

\subsection{Radio emission shaping the X-ray cavity}
\label{sec:discuss-hole}

The X-ray cavity is a region of distinctly reduced X-ray surface
brightness in the direction where the radio emissions from
\seight\ and \source\ overlap.  Due to coincidence in
  projection, we assume the cavity is co-located within the core
  radius of the cluster, and we have demonstrated that the contrast in
  X-ray surface brightness is consistent with such an interpretation.

Many clusters and groups are now known to harbor gas cavities
containing the radio plasma of active galaxies
\citep[e.g.,][]{birzan04}.  Since the pressure in relativistic plasma
should not be less than that of surrounding X-ray gas, insight can be
gained into the extent to which the relativistic plasma departs from a
state of minimum energy or requires extra pressure to be provided by
non-radiating particles or a reduced filling factor
\citep*[e.g.,][]{dunn, birzan, osullivan}. Where a detection of
inverse-Compton radiation from the cavities can be used to measure the
magnetic field, the minimum-energy assumption is tested, and results
concerning the other pressure components are more secure
\citep[e.g.,][]{wfos}.  In \cluster, while there is not a clean
spectral separation of inverse-Compton and thermal emission in the
cavity, the suggestion of lower metallicity implies that
inverse Compton emission is not completely negligible, providing a
modeling constraint.

The average gas pressure surrounding the X-ray cavity can be
calculated from the X-ray radial profile, giving\footnote{1 Pa = 10
  dynes cm$^{-2}$} $(9.8 \pm 0.5) \times 10^{-13}$ Pa ($1\sigma$
error).  A sphere of radius 100 arcsec evacuated of
  X-ray-emitting gas provides a model for
  the cavity that is consistent with the
surface-brightness distribution (Fig.~\ref{fig:cavityplotit}).  We model the
cavity's radio emission from our map using data from program AC105 as
synchrotron radiation from electrons with a power-law distribution of
$\alpha_{\rm r}=0.9$ extending to a minimum Lorentz factor of
$\gamma_{\rm min}=10$, to find a minimum-energy field strength
\footnote{1 nT = 10 $\mu$G} 
in the cavity of $B_{\rm me}=0.85$ nT (no protons) or 1.0 nT 
(equal energy in protons and electrons).  We don't
have radio data to constrain $\alpha_{\rm r}$, but the adopted value
is supported by the X-ray data if modeled as
power-law emission from inverse Compton scattering by the electron
population (\S \ref{sec:armhole}).  The resulting pressure is a factor
of 5.0 (3.5) too low to match that in surrounding X-ray gas, where
here and in what follows the first value is for no protons and the
second for equal energy in protons and electrons.

While the cavity pressure can be increased by adding further
non-radiating particles or reducing the filling factor, the prediction
for inverse Compton scattering on the cosmic microwave background
radiation is still then less than 6 per cent of the X-ray counts from
the cavity, not explaining the reduced metallicity when the spectrum
is fitted entirely to thermal emission.  In contrast, a small
departure from minimum energy in the sense of a reduced magnetic field
can both provide the required pressure and explain the metallicites.
Specifically, reducing the magnetic field strength to 30 (37) per cent
of the minimum-energy value, i.e., to 0.26 (0.37) nT, matches the
pressure and provides an increase in electron number density (to match
the radio synchrotron emission) such that 52 (27) per cent of the
0.3--5-keV X-ray counts from this region would be from inverse Compton
emission.  A modest departure from minimum energy is consistent with
findings elsewhere, and in particular with the lobes of radio galaxies
for which X-ray inverse Compton is detected and modeled using the same
value of $\gamma_{\rm min}$ as adopted here \citep{croston}.
We then note that the cavity shape would need to be somewhat
  spheroidal and elongated along the line of sight, to explain the
  surface brightness contrast of Figure~\ref{fig:cavityplotit} in the presence of extra
  (inverse Compton) X-ray counts in the cavity.

The sound speed in X-ray gas of temperature $kT$ (keV) is roughly $516
\sqrt{kT}$ km s$^{-1}$, meaning that radio plasma should have been
developing and filling the cavity for at least $10^8$ yrs for the
cavity not to have collapsed.  This is within the lifetime of typical
radio galaxies based on spectral aging measurements
\citep[e.g.][]{alexander}.  \seight, whose southern lobe is currently
at the edge of the cavity, would have taken a comparable time ($1.7
\times 10^8$ yr) to cross the cavity at a speed characteristic of the
cluster velocity dispersion, and faster speeds are expected here, near
the cluster center.  Thus it seems likely that the radio plasma of
\seight\ is primarily responsible for carving out the cavity.  In
contrast, the radio emission of \source\ is broken into the swirl
inside the cavity.  While this might be due to collision with radio
plasma of \seight, it may also result from interaction with a wake
left by the motion of \seight\ and its companion galaxy \citep[see
  e.g.,][]{wb346}.

If \seight\ is responsible for the cavity, its lobes should be at
least mildly overpressured with respect to the surrounding gas.  The S
lobe lies at the entrance to the cavity where the average gas pressure
(calculated using the X-ray radial profile for an off-center angular
distance of 65 arcsec) is $(1.11 \pm 0.05) \times 10^{-12}$ Pa
($1\sigma$ error).  We model the lobe using our 1.39-GHz JVLA data
with parameters as for the cavity\footnote{The visibilities in the
  C-band data miss too much of the larger-scale lobe structure to
  provide a good value for $\alpha_{\rm r}$.}, but with a sphere of
radius 18 arcsec, to find a minimum-energy field strength of $B_{\rm
  me}=2.2$ (2.6) nT and a pressure of $1.3 (1.8) \times 10^{-12}$ Pa,
slightly above that in the external gas.  If the magnetic field is
below the minimum-energy value by the amount we argue is likely for
the cavity, then the S lobe is overpressured by a factor of 4.9 with
respect to its surrounding gas.  Such an overpressure would lead to a
shock of Mach number 2 being driven into the external medium
\citep[e.g., equation 49 of][]{wbrev} and direct heating in the
immediate vicinity of the lobe, not ruled out by the data due to
masking by much foreground and background gas along the line of sight.
Direct evidence of shock heating is difficult to verify around
cluster-embedded radio lobes, but there is growing evidence for
moderately strong shocks around intermediate-power radio galaxies like
\seight\ \citep{wfos}.

\subsection{X-ray gas shaping radio features}
\label{sec:discuss-inner}

It is striking that the radio filament of the S lobe of
\seight\ (Fig~\ref{fig:radio}) runs along the northern inside edge of
the cavity (Fig~\ref{fig:large}).  The structure contains strands, and
the brightening is suggestive of preferential particle acceleration
along the interface between the cavity and the external X-ray gas ---
something which in principle could be tested by radio spectral index
mapping.

The tendrils of \seight\ run along the outside of the plateau and look
like they are buoyant, so that their deprojected pressures should
match those in the plateau, as measured using the spectral fit and
profile in Figure~\ref{fig:profilesEW}.  To test this we take
rectangular sections at the end of each tendril modeled as cylinders
of radius 21 arcsec and lengths 126 and 162 arcsec centered at
distances from the center of the gas distribution of 380 and 530
arcsec for the SW and NW tendrils, respectively.  If the tendrils are
in the plane of the sky, the external pressure in the SW and NW is
$5.0 \times 10^{-13}$ and $3.6 \times 10^{-13}$ Pa, respectively.  For
minimum energy the two tendrils then agree in being under-pressured by
a factor of 5.5 (3.9), where again the two values correspond to a
lepton-only plasma or one where the lepton energy density is matched
by that in protons.  If in the plane of the sky with no significant
entrainment, pressure balance can be restored if the magnetic field is
29 (35) percent of the minimum-energy value.  These departures from
minimum energy are remarkably similar to those required in the cavity,
and by the region in the NW tendril the magnetic field strength so
estimated would have dropped to about half that in the cavity.
However, while there may be reasons for dynamic structures to depart
from minimum energy, it is more appealing for these buoyant flows to
have reached a state of minimum energy.  The tendrils extend into the
region where the X-ray gas pressure is falling with radius, $r$, as
roughly $r^{-1.7}$, and they would be in pressure balance with this
gas if lying at about $\theta = 25 (29)$ degrees to the line of sight.
The actual value of $\theta$ is likely to be set by the unknown
direction of the velocity vector of \seight, such that the buoyant
tendrils trail, and such values of $\theta$ are plausible.  Some
increase in internal pressure in the tendrils through gas entrainment
is also likely, allowing $\theta$ to increase from these estimates.
Not being in the plane of the sky has the advantage of explaining the
rather dramatic fading of the tendrils which occurs at larger $r$ for
smaller $\theta$. Strong fading is expected sufficiently far out in
the cluster outer atmosphere that the tendrils are no longer
supported, and the relativistic plasma expands adiabatically.  This is
easily accommodated by a further steepening in the radial profiles
beyond the values of $r$ for which we are able to make measurements in
the current data.

The broad tendril to the SE is the extension of the brighter
(approaching) jet of \source.  The two jet bends, first to the south
and then to the east (Fig.~\ref{fig:radio}) and presumably the result
of shocks, are sufficiently large in projection to suggest that this
jet is at a relatively small angle to the line of sight.  We test a
cylinder of radius 33 arcsec and length 98 arcsec (position angle 70
degrees) at 340 arcsec from the center of the gas distribution where
the external pressure (in the cooler gas here) is $2.8 \times
10^{-13}$ Pa.  As for \seight's tendrils, there is an underpressure by
a factor of a few that can be restored if the radio emission is lying
at 24 (31) degrees to the line of sight.

\subsection{Merging and gas flows}
\label{sec:discuss-outer}

X-ray cavities are sufficiently common that they are regarded as an
important heat source, with enough power to balance radiative cooling
in dense cluster cores \citep[e.g.,][]{dunn, rafferty}.  In the case
of the \cluster's cavity, we can estimate the change in enthalpy as
$\gamma PV/(\gamma-1)$, where $P$ is the pressure of the X-ray gas,
$V$ is the volume of the cavity, and $\gamma = 4/3$ is the ratio of
specific heats for relativistic gas.  The resulting value is roughly
$2 \times 10^{53}$ J.  In \S \ref{sec:integrated} we pointed
out that in its integrated properties the cluster is too hot by
$k\Delta T \approx 1.5$ keV for its luminosity, based on scaling
relations.  This is equivalent to an excess enthalpy of $\gamma M_{\rm
  gas}k\Delta T/\mu m_{\rm H}(\gamma-1)$, where $M_{\rm gas}$ is the
total gas mass, $\mu m_{\rm H}$ is the particle mass, and $\gamma =
5/3$ is the ratio of specific heats for the X-ray-emitting gas.  The
resulting value is roughly $1.7 \times 10^{55}$ J.  A
staggering 85 cavities like those attributed to
\seight\ would be needed to explain the excess heat in \cluster\ that
causes it to deviate from scaling relations.

The most likely source of the excess heat is therefore a merger.  The
temperature structure seen in Figure~\ref{fig:contbin} points to such
a merger being roughly along a NW-SE direction, such that gas heated
in collisions escapes to the sides.  It is also consistent with the
velocity data discussed in \S\ref{sec:velocities} and the two centers
marked as crosses in Figure \ref{fig:armhole}.  If we take the merging
group to have about 5 per cent of the mass of \cluster, based on the
number of galaxies within known velocities, and use the
radial-velocity offset of roughly 1700 km s$^{-1}$ as the total
velocity difference, then the conversion of 20 per cent of the kinetic
energy of the group would be sufficient to produce the required excess
heating. That is, the heating requires only a minor merger event. The
absence of a shock feature in the X-ray image is not unexpected if
this merger is indeed responsible for the heating, since such shocks
are only seen when the merger is in the plane of the sky, and the
radial velocity difference would be excessive if the merger axis is
far from the line of sight in the present case.

Numerical simulations have shown that mergers lead to persistent
(gigayears) relative motion of gas streams which are in pressure
equilibrium (sloshing), and this can explain X-ray features seen
rather commonly in relaxed cool-core galaxy clusters and known as cold
fronts, where a large density discontinuity is in pressure balance due
to the dense gas being cooler \citep{yago, mv}.  \cluster\ is far from
relaxed, and shows no obvious evidence of sloshing. We clearly see
density substructure, but the temperature and density features lie in
different parts of the gas, the former having a closer relationship to
the location of the radio tendrils.  The most likely explanation is
that we are witnessing a relatively recent merger encounter, one which
may also have helped trigger the radio galaxies into their current
phase of activity.

\subsection{Sliding and lubrication}
\label{sec:discuss-lubrication}

The excess X-ray emission in the region we name the `plateau' can be
understood since the centroid of the distribution of galaxies with
reported velocities (biased towards the more massive ones) lies
in this neighborhood.  Its temperature is unremarkable
compared with that in the X-ray brightest region, including
the temperatures in the arm and surrounding the cavity.  What
is remarkable is that the radio tendrils of \seight\ appear to
  border the hottest gas (Fig.~\ref{fig:contbin}), and envelop the
  roughly cylindrical gas region seen in Figure~\ref{fig:armhole} that
  forms the bright X-ray bridge between the arm and the plateau.
It is tempting to invoke cause and effect, such that the radio plasma
of \seight\ helps to reduce thermal transport between the
  warmer and cooler gas, and to reduce momentum transport and hence
  viscous drag, so that X-ray gases of different temperatures can
slide by one another with little heat exchange or mixing.
The regions in the temperature map were defined automatically from the
smoothed X-ray data, with no reference to the radio structures. It is
also notable that the southern tendril of \source\ lies along a
temperature boundary.

The morphology of the radio and X-ray extensions suggest that
  the radio extensions of \seight's lobes were caused by the relative
motion of the host galaxy and the gas extending into the
  plateau.  The attachment of radio plasma to this infalling gas would
  cause the extension of the radio lobes into the tendrils and cause
  the gas bridge to become at least partially enveloped by the radio
plasma. Because the radio plasma is moderately strongly magnetized, it
will act as an effective barrier to transport between the inner arm
and the general cluster environment. Not only are the particle
gyroradii smaller (by a factor of about 10) in the radio plasma than
in the general cluster environment, but the field will also tend to be
ordered by the stretching along the radio tendrils. This will reduce
the (perpendicular) thermal conductivity between the two gas regions.

The degree of thermal protection offered by the radio features depends
both on the reduction in transport through the radio plasma and the
fraction of the interface between regions of different temperature
covered by the radio plasma. Taking Spitzer conductivity 
$$\kappa \approx 5 \times 10^{-12} (T/{\rm
  K})^{5/2}~{\rm~W~m}^{-1}~{\rm K}^{-1}$$
\citep{spitzer} appropriate for intracluster gas, and modeling
  the gas as a cylinder of radius $R \approx 50$~kpc, the timescale for
  a temperature difference $\Delta T$ to be erased is of the order
$$ \tau \approx {n_o R^2 k \over \kappa} \left({T \over \Delta
  T}\right)$$ or about 100 Myr, taking the proton density,
  $n_o$, to be roughly 940 m$^{-3}$ (\S \ref{sec:integrated}). This is significantly shorter
  than the time (about 300 Myr) taken for the gas to travel from
  \seight\ to the plateau at the sound speed.  However, it appears
  that at least half the surface area is covered by radio plasma.  If
  this radio plasma cuts $\kappa$ by a factor of about 10, then
  effective heat
  transport occurs over only about half the surface of the cylinder
  and its thermal lifetime is raised by a factor of order 3.  The
  associated drop in viscosity may also help the gas to slide in
  towards the plateau.

\subsection{The X-ray hotspots of \seight}
\label{sec:discuss-hotspots}

There is good evidence of X-ray emission from both the east and west
components of the northern double hotspot of \seight, each with
similar ratios of X-ray to radio flux density.  The X-ray emission is
about two orders of magnitude higher than expected due to
inverse-Compton scattering if the electrons and magnetic field are at
minimum energy.  This result is in common with many other
\chandra-observed hotspots, particularly those in FR\ II radio
galaxies at the lower end of the spectrum of total radio power
\citep{hard-hot}, The large reduction in magnetic field (about a
factor of 24 for \seight) and the subsequent high increase in source
energy needed to explain the X-ray emission by the inverse-Compton
mechanism led \citet{hard-hot} to suggest that the electron spectrum
extends to high enough energies in such hotspots for synchrotron
radiation to dominate the X-ray output.  In \seight's hotspots, if the
electron spectrum breaks from a power law slope of $p=2.44$ (giving
the observed $\alpha_{\rm r} = 0.72$) by $\Delta\alpha$=1 at a Lorentz
factor of $\gamma = 6 \times 10^4$, the synchrotron radiation from
100-TeV electrons could be responsible for the X-rays in both
components in minimum-energy magnetic fields of about 10 nT.  While
this is an attractively simple explanation, we note that
\citet{werner} have used \spitzer\ data to claim that a
single-component broken-power-law synchrotron spectrum can be excluded
for 80 per cent of 24 hotspots they study.  They point to the work of
\citet{deyoung} as possible support for magnetic fields in hotspots
not having increased to minimum-energy levels.

An alternative reason that hotspots on the jet side of a source (as in
\seight) may be unusually X-ray bright is if relativistic beaming is a
factor.  An applicable mechanism suggested by \citet{georg} is that
for a jet approaching a hotspot with high bulk Lorentz factor the
available photon fields for Compton scattering will be boosted by the
strongly directional (particularly in the jet frame) radio synchrotron
emission from the terminal hotspot, resulting in inverse Compton
X-rays beamed in the forward direction of the jet and offset
(upstream) from the peak of the radio emission.  At high redshift the
spatial separation of components necessary to test this mechanism is
not possible, although a claim at low redshift has been made for PKS
B2152-699 \citep{wfos}.  In \seight\ there is some evidence for an
offset between radio and X-ray emission, although we cannot rule out
residual uncertainties in our 5-GHz radio mapping 3.5 arcmin off axis
as responsible.  A non-detection of the southern hotspot region, at
the termination of the jet pointed away from the observer, is
consistent with beaming playing a part in the detection of the X-ray
hotspots, although statistics are such that the formal upper limit on
the X-ray flux is at a similar level to the detections in the north,
and so without more sensitive X-ray data firm conclusions cannot be
drawn.

\section{Summary}
\label{sec:summary}

We have presented results using new \chandra\ and JVLA observations of
\cluster, together with archival lower-resolution VLA data.  We find
the cluster to be far from relaxed, and devoid of a cool core.  The
gas is too hot on average by $k\Delta T \approx 1.5$ keV for the
cluster to agree with temperature-luminosity correlations, and while
there is a 100 kpc-scale cavity carved out by radio-emitting plasma,
the excess enthalpy is insufficient to explain the heating.

Much of what is observed seems likely to have been caused by the
recent merger of a small subgroup of galaxies with fast relative
motion at relatively small angle to the line of sight in a roughly
SE-NW projected direction.  This can easily produce enough heating,
and seems to explain the temperature distribution, where hotter gas
lies in directions perpendicular to the inferred line of encounter.
Existing galaxy-velocity data provide some support for this
minor-merger hypothesis, but a more detailed survey of galaxy
velocities in the field is needed to provide a more complete test.

While the X-ray data provide no evidence for shocks (understandable
since velocity data suggest the encounter is not in the plane of the
sky), or isobaric density drops (cold fronts) that would be
explainable by gas sloshing, there is much interplay between the X-ray
gas and the relativistic plasma hosted by \source\ and \seight.  This
radio-emitting plasma terminates in buoyant tendrils reaching the
cluster's extremities.  In the case of \seight, a tendril trails from
each radio lobe, almost certainly as a result of the galaxy's motion
and consequent drag.  In contrast, the counter jet of \source\ runs
into the X-ray cavity, where it produces a dramatic swirl perhaps due
to collision with the radio plasma from \seight\ or in the wake of its
motion, but a tendril extends from the highly bent southeastern jet.
An important and noticeable feature of the tendrils is that they run
along boundaries between gas of different temperatures.  Because the
radio plasma is moderately strongly magnetized, it will act as an
effective barrier to transport between gas layers and reduce the
effective viscosity, helping to preserve post-merger gas flows.

The radio galaxies hosted by \source\ and \seight\ both have powers in
the range that dominates output in the Universe as a whole, and so are
most important for understanding radio-mode feedback.  The most
apparent ways in which this feedback is mediated by these sources is
through the large X-ray cavity and the capacity for the tendrils to
allow the sliding of gas flows and the containment of temperature
structures.  In other respects the X-ray properties of the radio
galaxies are normal for their radio powers.  The power-law X-ray
emission from both nuclei is unabsorbed and lies within the scatter of
radio/X-ray correlations, arguing in favor of a common non-thermal
origin for the radio and X-ray core emission.  Resolved X-ray emission
is detected from the brighter jet in each source.  X-rays are detected
from both components of the northern double hotspot of \seight, which
we have discussed in some detail given the on-going debate concerning
the flow-speed into hotspots and the origin of their X-ray emission.

\cluster\ provides a particularly interesting case study of plasma
dynamics, minor-merging, and radio-galaxy feedback in a non-cool core
cluster, worthy of deeper study.

\acknowledgments

We acknowledge support from NASA grant GO1-12010X.  
We thank the anonymous referee for constructive comments, and
Paul Giles for discussion of the temperature luminosity relation.
Results are
largely based on observations with \chandra, supported by the CXC.
The National Radio Astronomy Observatory is a facility of the National
Science Foundation operated under cooperative agreement by Associated
Universities, Inc.  This research has made use of the NASA/IPAC
Extragalactic Database (NED) which is operated by the Jet Propulsion
Laboratory, California Institute of Technology, under contract with
the National Aeronautics and Space Administration.  This publication
makes use of data products from the Two Micron All Sky Survey, which
is a joint project of the University of Massachusetts and the Infrared
Processing and Analysis Center, funded by the National Aeronautics and
Space Administration and the National Science Foundation.

Facilities: \facility{CXO, VLA}

\end{document}